\documentclass[prb,aps,english,twocolumn,notitlepage,superscriptaddress]{revtex4-2}
\usepackage{epsfig}
\usepackage{amsmath}
\usepackage{mathtools}
\usepackage{xcolor}

\DeclarePairedDelimiterX\braket[2]{\langle}{\rangle}{#1 \delimsize\vert #2}
\usepackage{sidecap}
\usepackage[figuresleft]{rotating}
\usepackage{caption}
\usepackage{float}
\usepackage{appendix}
\usepackage{subfigure}
\usepackage{graphicx}
\usepackage{hyperref}
\usepackage{array,multirow}
\usepackage{multirow}
\usepackage{physics}
\usepackage{tabularx}
\usepackage{natbib}
\usepackage{comment}

\usepackage[normalem]{ulem}

\begin{document}
\title{Identification of odd-frequency superconducting pairing in Josephson junctions}

\author{Subhajit Pal}
\email{subhajitp778@gmail.com, \\ ORCID ID :   
0000-0002-2419-1337}
\affiliation{Indian Institute of Science Education \& Research Kolkata,
Mohanpur, Nadia - 741 246, 
West Bengal, India}

\author{Aabir Mukhopadhyay}
\email{aabir.riku@gmail.com, \\ ORCID ID : 0000-0001-6465-2727}
\affiliation{Indian Institute of Science Education \& Research Kolkata,
Mohanpur, Nadia - 741 246, 
West Bengal, India}

\author{Vivekananda Adak}
\email{adakvivek0705@gmail.com, \\ORCID ID: 0000-0002-0723-4155}
\affiliation{Department of Physics, Korea Advanced Institute of Science and Technology,
Daejeon 34141, Korea}

\author{Sourin Das}
\email{sourin@iiserkol.ac.in, \\ ORCID ID : 0000-0002-8511-5709}
\affiliation{Indian Institute of Science Education \& Research Kolkata,
Mohanpur, Nadia - 741 246, 
West Bengal, India}
\pacs{74.50.+r,74.78.Na,85.25.-j,85.25.Cp,85.65.+h,75.50.Xx,85.80.Fi}
\begin{abstract}
{Optimal choice of spin polarization enables electron injection into the helical edge state at a precise position, despite the uncertainty principle, permitting access to specific non-local Green's functions. We show, within 1D effective description, that this fact facilitates a direct identification of odd-frequency pairing through parity measurement (under frequency reversal) of the non-local differential conductance in a setup comprising Josephson junction on the helical edge state of a 2D topological insulator with two spin-polarized probes tunnel-coupled to the junction region. A 2D numerical simulation has also been conducted to confirm theoretical predictions as well as to demonstrate the experimental feasibility of the proposal.}
\end{abstract}
 \maketitle
\textit{Introduction:} 
In a conventional superconductor, electrons form Cooper pairs due to a weak attraction mediated via lattice vibrations\cite{bard}. The pairing function between the two electrons within a Cooper pair can be classified into different symmetry classes by means of four permutation operators with respect to spin ($S$), relative coordinate ($P^*$), orbital index ($O$), and time coordinate ($T^*$), all of which can only have eigenvalues $\pm 1$.
Generally, BCS theory of single orbital superconductor ($\expval{O}=1$) is instantaneous in time ($\expval{T^*}=1$). This leaves two choices for the combinations of eigenvalues of the operators $S$ and $P^*$ to satisfy the Berezinskii condition\cite{bere} $SP^*OT^*=-1$ for the pairing function, i.e., $\{ 
\expval{S}, \expval{P^*}\}=\{ -1,1\}, \{1,-1\}$. These correspond to spin-singlet even-parity (e.g. $s$-wave) and spin-triplet odd-parity\cite{sig} (e.g. $p$-wave) symmetries, respectively.
If one removes the constraint of instantaneous pairing, the possibility of temporal (non-local in time) Cooper pairing becomes feasible, and both $\expval{T^*}=\pm 1$ can be possible, depending on whether the pairing function is symmetric or anti-symmetric under the permutation of relative time coordinates of the electrons forming the Cooper pair. Temporally symmetric pairing ($\expval{T^*}=1$) does not give rise to any new symmetry classes other than those already described by BCS theory and is generally known as even-frequency (even-$\omega$) pairing. However, temporally anti-symmetric pairing ($\expval{T^*}=-1$), which is known as odd-frequency (odd-$\omega$) pairing\cite{lin,tana,cay}, opens up possibilities of two new symmetry classes (assuming $\expval{O}=1$) that cannot be described by BCS theory, namely, spin-singlet odd-parity ($\{ 
\expval{S}, \expval{P^*}\}=\{ -1,-1\}$) and spin-triplet even-parity ($\{ 
\expval{S}, \expval{P^*}\}=\{1,1\}$) pairings, consistent with Berezinskii condition. It can also be shown that this odd-$\omega$ (even-$\omega$) pairing is also odd (even) under the sign change of energy\cite{lin}, which gives the advantage of observing the behavior of this pairing in the energy domain rather than in the temporal domain.

On a historical note, the concept of odd-$\omega$ spin-triplet pairing traces back to Berezinskii's 1974 proposal in $^3$He\cite{bere} and predictions in disordered systems\cite{trk,bel}. Balatsky and Abrahams later indicated the presence of odd-$\omega$ spin-singlet pairing in time-reversal and parity symmetry broken superconductors\cite{bala}. Subsequently, the odd-$\omega$ pairing was explored within the framework of a two-channel Kondo system\cite{eme}, the 1D $t-J-h$ model\cite{bon}, the 2D Hubbard model\cite{bulu} and heavy fermion compounds\cite{cole}. The bulk odd-$\omega$ pairing has been indirectly hinted theoretically through the Majorana scanning tunneling microscope\cite{oka} and also experimentally using the Kerr effect\cite{ers,kom}, and the paramagnetic Meissner effect\cite{adi,bse,ali,jak,yta}. 
Further, the evidence of bulk odd-$\omega$ pairing due to magnetic impurities has emerged in $s$-wave superconductors\cite{perr,kuzz}. In Ref.~\cite{vko}, a measurement tool was introduced for directly detecting odd-$\omega$ pairing in bulk systems using time- and angle-resolved photoelectron fluctuation spectroscopy. However, this method requires advanced technology beyond the current capabilities of facilities. More recently, in Ref.~\cite{dca}, authors proposed a detection scheme to directly identify odd-$\omega$ pairing in bulk systems using the quasiparticle interference method in the presence of an external magnetic field. Initially, odd-$\omega$ pairing was regarded as an inherent bulk phenomenon\cite{bere,bala,eab} but was subsequently acknowledged to manifest in heterojunctions\cite{berg1,berg2,berg3,kopu,vol,toko,tan,amb1,amb2,cre,burr,cayao,hwa,cay1,lind1,lindd2,pal,tamu,aag,lof,vol1,fom,buz,ats,asan,dku,CFL,tanaka2021theory,PhysRevLett.99.037005} and in the systems under the influence of time-dependent fields\cite{trio1,trio2}. Certain theoretical studies have indirectly identified odd-$\omega$ pairing in heterostructures. For instance, this was achieved through phase-tunable electron transport in topological Josephson junctions (JJ)\cite{jca}, examining Josephson current characteristics on the surface of Weyl nodal loop semimetals\cite{pdut}, and analyzing current noise in JJ\cite{rse}. Experimental evidence of odd-$\omega$ pairing in heterostructures emerged through measurements of long-range supercurrents in magnetic JJ\cite{kha,adb}.
These two experiments utilize a ferromagnetic material to create an odd-$\omega$ pairing effect, which is subsequently identified by measuring long-range superconducting pairing. In contrast, this letter investigates a configuration that inherently supports odd-$\omega$ pairing without the need for any ferromagnet/superconductor junction. Topological JJ at the edge of a two-dimensional (2D) quantum-spin-Hall insulator (QSHI) serves this purpose. {It is important to mention here that induced superconductivity in the QSHI edge has already been experimentally demonstrated\cite{yacoby_NP_10_638}.}

We demonstrate that it is possible to read off all the independent, spatially non-local Green's functions by measuring differential conductance between {two} spin-polarized {probes} placed at the junction. {The underlying interplay between spin-momentum locking of helical edge states (HES) and spin-polarization of the probes allows for this advantage. Experimentally, the injection of spin-polarized current into the HES of QSHI has already been achieved\cite{molenkamp_NP_8_485}.} By using spin-polarized probes, one can inject {(detect)} electrons into {(from)} the helical edge with well-defined position and momentum simultaneously, which, in general, is prohibited due to uncertainty principle\cite{suma}.To elaborate further, a spin-$\uparrow$ {(spin-$\downarrow$)} electron injected at position $x$ in the helical edge can only tunnel into the right (left) moving edge mode owing to spin-momentum locking.
Keeping the spin-polarization axes of the tunneling probes aligned parallel or anti-parallel to the spin-quantization axis of HES, we study the difference between non-local differential conductance from the left probe to right probe $\kappa^{21}_{s\bar{s}}(V,0)$ and from right probe to left probe $\kappa^{12}_{\bar{s}s}(0,V)$ i.e.
\begin{align}
    \mathcal{A}^{2,1}_{s,\bar{s}}   
    &=\kappa^{21}_{s\bar{s}}(V_1=V,V_2=0)-\kappa^{12}_{\bar{s}s}(V_1=0,V_2=V)   \nonumber \\
    &=\left[\dfrac{dI_{s\bar{s}}^{21}}{dV}\right]_{(V_1=V,V_2=0)}-\left[\dfrac{dI_{\bar{s}s}^{12}}{dV}\right]_{(V_1=0,V_2=V)}    \label{Eq1}
\end{align}
where $\{ s,\bar{s}\}\in \{ \uparrow, \downarrow \}$ with $s\neq \bar{s}$, and $V_1(V_2)$ denotes the applied voltage at the left (right) tunneling probe {$P_1$ ($P_2$)}. It can be shown that this quantity $\mathcal{A}^{2,1}_{s,\bar{s}}$ can be written as a product of odd-$\omega$ and even-$\omega$ pairing amplitudes and hence should be odd with respect to incident energy {(and hence with respect to the applied voltage $V$)}.
We show that this difference stems from the same origin as that of the odd-$\omega$ part of the non-local differential conductance, which can be accessed via bias reversal i.e. $\mathcal{B}^{2,1}_{s,\bar{s}}=\kappa^{21}_{s\bar{s}}(V,0)-\kappa^{21}_{s\bar{s}}(-V,0)$. This fact leads to 
\begin{equation}
\mathcal{A}^{2,1}_{s,\bar{s}}=\mathcal{B}^{2,1}_{s,\bar{s}} \label{Eq2}.
\end{equation}
From an experimental perspective, measuring the anti-symmetric behavior of each of these independent quantities and their equality confirms the existence of odd-$\omega$ pairing. {In what follows, we will first present our results using an effective 1D analytic model of HES. Following that, we'll provide a 2D simulation of a realistic setup, in line with HgTe/CdTe quantum well parameters, confirming our predictions.}

\begin{figure}[t]
\centering
\includegraphics[width=0.48\textwidth]{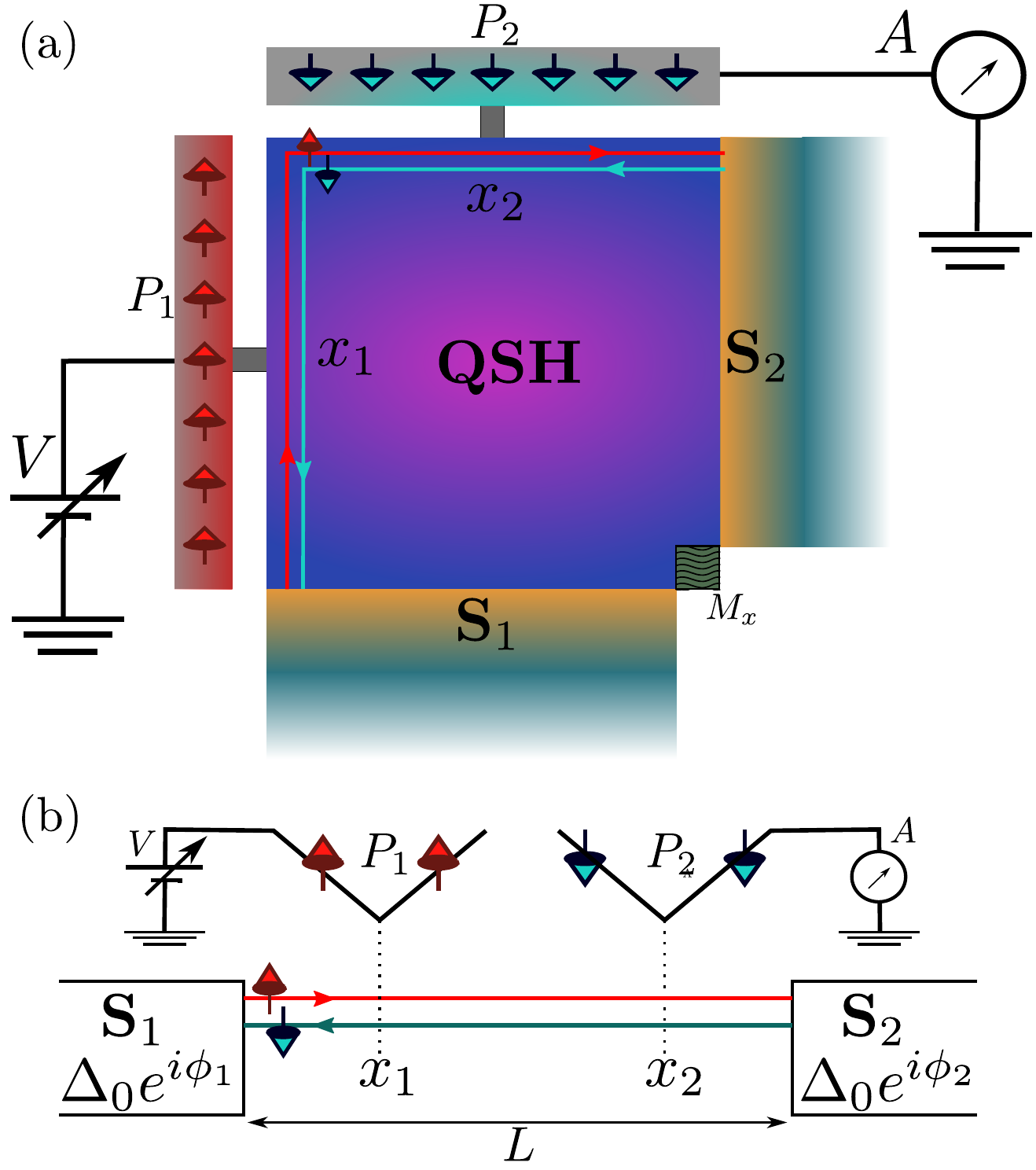}
\caption{\small \sl {(a) Schematic of the setup consisting of a Josephson junction at the edge of a 2D QSHI with two spin-polarized tunneling probes ($P_{1}$ and $P_{2}$) at $x=x_{1}$ and $x=x_{2}$. Up and down arrows represent the spin polarization direction of helical edge states as well as of the spin-polarized tunneling probes. A bias voltage $V$ is applied to $P_{1}$ while the current through $P_2$ is measured by the current read-out device $A$. (b) A simplified cartoon representation of the above setup.}}
\label{device}
\end{figure}

\textit{Model and its motivation:}
The system, we are interested in, is a JJ of length $L$ ($0\leq x\leq L$), realized in a helical edge state (HES) of a 2D QSHI\cite{lu,lu1,clk,call,keidel_PRR_2_022019} (1D Dirac fermions) which is proximitized to a conventional $s$-wave superconductor. {Schematic of the setup is shown in FIG. \ref{device} (a), while FIG. \ref{device} (b) shows the corresponding simplified cartoon representation of the same setup.} The Bogoliubov-de Gennes (BdG) Hamiltonian characterizing such junctions can be written as\cite{aab1,aab2} $H_J=\int_{-\infty}^{\infty} dx\, \Psi^{\dagger} \mathcal{H}_J \Psi$ where $\Psi=(\psi_{\uparrow}, \psi_{\downarrow}, \psi_{\downarrow}^{\dagger}, -\psi_{\uparrow}^{\dagger} )^{T}$ is the Nambu basis and
\begin{equation}
    \mathcal{H}_{J}= -i\hbar v_{F}\partial_{x}\tau_{z}\sigma_{z}-\mu\tau_{z}+{\Delta}(x)(\cos{\varphi_r}\tau_{x}-\sin{\varphi_r}\tau_y)\,.
    \label{Ham_J}
\end{equation}

Here, $\sigma_i$ and $\tau_i$ are the Pauli matrices acting respectively on the spin basis and particle-hole basis, $\mu$ is the chemical potential of the HES, $v_F$ {represents} Fermi velocity of the HES, and, {$\hat{p}=-i\hbar\partial_x$ is the momentum operator.} The spin-quantization axis of the HES is considered to be along $z$-direction\cite{aab1,aab2}. The superconducting pairing potential is given by $\Delta(x)=\Delta_0[ \Theta(-x)+\Theta(x-L)]$. {Superconducting leads ($S_r$) are identified as $r \in \{1,2\}$ with the corresponding superconducting phases $\varphi_{r}$ such that the superconducting phase difference $\varphi_{21}=\varphi_{2}-\varphi_{1} \neq 0$, in general.}

\begin{table*}[ht]
    \centering
    \begin{tabular}{|c|c|c|c|c||c|c|c|c|c|}
        \hline
        Input &Polarization & Polarization    &Output &Contributing &Input &Polarization & Polarization    &Output &Contributing   \\
        at $P_1$    &of $P_1$($\theta_1$)  &of $P_2$($\theta_2$) &at $P_2$  &Green's function   &at $P_2$    &of $P_2$($\theta_2$)  &of $P_1$($\theta_1$) &at $P_1$  &Green's function   \\
        \hline
        e   &$\uparrow$     &$\uparrow$    &e  &$G^{r(a)}_{ee,\uparrow\uparrow}(x_2>x_1)$    &e   &$\uparrow$     &$\uparrow$    &e  &$G^{r(a)}_{ee,\uparrow\uparrow}(x_1<x_2)$ \\
        \hline
        e   &$\uparrow$  &$\downarrow$   &h  &$G^{r(a)}_{he,\downarrow \uparrow}(x_2,x_1)$   &e   &$\uparrow$  &$\downarrow$   &h  &$G^{r(a)}_{he,\downarrow \uparrow}(x_1,x_2)$  \\
        \hline
        e   &$\downarrow$    &$\uparrow$ &h  &$G^{r(a)}_{he,\uparrow \downarrow}(x_2,x_1)$   &e   &$\downarrow$    &$\uparrow$ &h  &$G^{r(a)}_{he,\uparrow \downarrow}(x_1,x_2)$  \\
        \hline
        e   &$\downarrow$    &$\downarrow$   &e  &$G^{r(a)}_{ee,\downarrow \downarrow}(x_2>x_1)$ &e   &$\downarrow$    &$\downarrow$   &e  &$G^{r(a)}_{ee,\downarrow \downarrow}(x_1<x_2)$    \\
        \hline
    \end{tabular}
    \caption{Output at $P_2$ ($P_1$) when an electron is injected to the HES through $P_1$ ($P_2$) and the corresponding Green's functions depending on the polarization of the tunneling probes}
    \label{advantage-table_1}
\end{table*}

Two spin-polarized tunneling probes\cite{wad} $P_{\{1,2\}}$ ($P_r$ being closer to $S_r$) with spin-polarization axes oriented respectively along $\hat{n}_1$ and $\hat{n}_2$, at angles $\theta_1$ and $\theta_2$ with respect to the spin-quantization axis of the HES (say, in the $z-x$ plane)\footnote{The azimuthal angle is of no consequence, e.g., it can also be assumed in the $z-y$ plane}, are introduced at positions $x=x_1$ and $x=x_2$ respectively within the junction region ($0< x_1< x_2 < L$). {For most of the paper, we will be only discussing the case for $\{\theta_1, \theta_2\}=\{0,\pi\}$, unless otherwise stated.} To keep the algebra simple while retaining the essential physical inputs, we model the probes as a spin-polarized 1D mode. The second quantized Hamiltonians of these probes can be expressed as:
\begin{equation}
    H_{P_{1(2)}}=-i \hbar v_F \int_{-\infty}^{\infty} d\Tilde{x}\, (\psi_{\hat{n}_{1(2)}}^\dagger \partial_{\Tilde{x}}\psi_{\hat{n}_{1(2)}}).
    \label{Ham_P}
\end{equation}

The tunneling Hamiltonian between the HES and the probes can be written as:
\begin{equation}
    H_{T}^{P_i} = \hbar v_F \int_{-\infty}^{\infty} {dx\, \delta(x-x_i)} \left( \sum_{\substack{\alpha, \alpha'\\ \alpha \neq \alpha'}} t^{i}_{\alpha \alpha^{'}} \psi_{\alpha}^{\dagger} \psi_{\alpha'} +\text{H.c.}\right)
    \label{tunnelling_ham}
\end{equation}
with $\alpha, \alpha' \in \{ \uparrow, \downarrow, \hat{n}_i \}$ {($\alpha\neq\alpha'$ and $i \in \{ 1,2 \}$)}. $t^{i}_{\alpha \alpha'}=t_i\gamma_{\alpha \alpha'}$ is the tunneling strength between $\alpha$ and $\alpha'$ at $x=x_i$ and $\gamma_{\alpha \alpha'}$ is the overlap of the spinor part of the first quantized wave function of electrons in the probes and the HES. {From this point onward, we shall consider $\hbar=v_F =1$.} Note that this form of tunneling respects $SU(2)$ symmetry and hence cannot induce spin-flip scattering {when $n_i$ is parallel or anti-parallel to the spin-quantization axis of the HES}. Using the equation of motion approach for the Hamiltonian $H=H_{\text{HES}}+\sum_{i\in \{1,2\}}H_{P_i}+H_T^{P_i}$ (where $H_{\text{HES}}=H_J|_{\Delta=0}$), the scattering matrices can be expressed as,
\begin{equation}
    \psi_{\alpha}^{\text{out}} (x_i) = \sum_{\alpha'} S^e_{\alpha \alpha'} (x_i) \psi_{\alpha'}^{\text{in}}(x_i) \, .
    \label{scattering_mat}
\end{equation}
where $\psi^{\text{in(out)}}$ are the corresponding incoming (outgoing) plane wave amplitudes on the HES and the probe. The scattering matrix for holes ($S^h$) can be determined by exploiting the particle-hole symmetry (for details, see Supplemental Material (SM) Section A) of the system.  
We are interested in studying charge transport between the two probes via the junction when a finite voltage bias $V$ is applied such that $-\Delta_0 \leq eV \leq \Delta_0$.

Note that, the spatial non-locality of the probes, along with the helical nature of the edge state, opens up the possibility of detecting only a hole at $P_2$ when an electron is injected into HES through $P_1$ provided the polarization of the probes is tuned such that $ \theta_1, \theta_2 =0 \text{ or }\pi$ and $\theta_1\neq \theta_2$. This leads to the fact that the differential conductance between the probes is directly proportional to the square of the modulus of anomalous Green's functions. 
This can be understood in terms of a simple process, as described below. If the spin polarization of $P_{1}$ is set to spin-$\uparrow$ (i.e., $\theta_1=0$), it can only inject a right-moving electron to the HES due to the spin-momentum locking of the edge states. In the absence of spin-flip scattering within the HES, this electron will either remain as a spin-$\uparrow$ electron after an even number of Andreev reflections (including the possibility of no Andreev reflection) or will convert to a spin-$\downarrow$ hole after an odd number of Andreev reflections when it reaches $x=x_2$. Thus, $P_2$ will detect only a hole for $\theta_2=\pi$. Along the same line of argument, it is straightforward to see that for $\theta_2=0$, $P_2$ will detect only an electron. By definition, the probability amplitude of these non-local transmissions will be proportional to the corresponding anomalous Green's functions, namely $t_{ee,\uparrow\uparrow}^{21} \propto G^{r(a)}_{ee,\uparrow\uparrow}(x_2>x_1)$ and $t_{he,\downarrow\uparrow}^{21} \propto G^{r(a)}_{he,\downarrow\uparrow}(x_2,x_1)$, where $t_{ee,\uparrow\uparrow}^{21}$ and $t_{he,\downarrow\uparrow}^{21}$ are the {propagation amplitudes for spin up electron and spin down hole respectively from the left probe to the right probe} and, $G^{r(a)}_{ee,s_2s_1}(x_2>x_1)$ represents the retarded (advanced) normal Green's functions while $G^{r(a)}_{he,s_2s_1}(x_2,x_1)$ ($s_i\in\{\uparrow,\downarrow\}$) denotes the retarded (advanced) anomalous Green's functions calculated at $x=x_{2}$. Here retarded and advanced Green's functions are defined in the appropriate frequency domain\cite{cayao}. The above discussion is summarized in the form of TABLE.~\ref{advantage-table_1}.

Using Landauer's formula\cite{glaz}, we can calculate the differential conductance from $P_1$ to $P_2$, given by $\kappa_{\downarrow\uparrow}^{21}=\kappa_{ee,\downarrow\uparrow}^{21}-\kappa_{he,\downarrow\uparrow}^{21}$ and from $P_2$ to $P_1$, given by $\kappa_{\uparrow\downarrow}^{12}=\kappa_{ee,\uparrow\downarrow}^{12}-\kappa_{he,\uparrow\downarrow}^{12}$, where $\kappa_{p_ip_j,s_is_j}^{ij}=(e^2/h) |t_{p_ip_j,s_is_j}^{ij}|^2$ ($p_{i(j)} \in \{ e,h \}$, $s_{i(j)}\in \{ \uparrow, \downarrow \}$, $\{i,j\}\in\{1,2\}$). The difference between these two non-local differential conductances at energy (frequency) $\omega$ is given by:
\begin{align}
     \mathcal{A}^{2,1}_{\downarrow,\uparrow} &= \kappa_{\downarrow\uparrow}^{21}- \kappa_{\uparrow\downarrow}^{12}
    =  -\kappa_{he,\downarrow\uparrow}^{21}+\kappa_{he,\uparrow\downarrow}^{12}, \nonumber \\
    \propto &-|G^{r}_{he,\downarrow\uparrow}(x_2,x_1,\omega)|^2 +|G^{r}_{he,\uparrow\downarrow}(x_1,x_2,\omega)|^2,   \nonumber   \\
    \propto &-|G^r_{he,\downarrow\uparrow}(x_2,x_1,\omega)|^2 +|G^a_{he,\downarrow\uparrow}(x_2,x_1,-\omega)|^2,   \nonumber  \\
    \propto &~ \text{Re}[\,\mathcal{O}[G^r_{he,\downarrow\uparrow}(x_2,x_1,\omega)] \times \mathcal{E}[G^{r*}_{he,\downarrow\uparrow}(x_2,x_1,\omega)] \,] 
    \label{anti-symmetry_in_I}.
\end{align}
where we have used the property $G^r_{he,s\bar{s}}(x,x',\omega)=G^a_{he,\bar{s}s}(x',x,-\omega)$ ($\{ s,\bar{s} \}\in \{ \uparrow,\downarrow \}$ and $s\neq\bar{s}$). Here we have used only retarded Green's functions to express $\kappa$. However, a proof involving advanced Green's functions follows in a similar way (See SM Section B). {In} Eq.~\eqref{anti-symmetry_in_I}, {$\mathcal{A}^{2,1}_{\downarrow,\uparrow}$} is \textit{anti-symmetric} with respect to $\omega$, and if measured, it can be treated as a direct signature of odd-$\omega$ pairing subjected to the condition that even-$\omega$ pairing within the junction is non-zero. In Eq.~\eqref{anti-symmetry_in_I}, $\mathcal{O}$ and $\mathcal{E}$ denote the odd-in-$\omega$ {(anti-symmetric with respect to $\omega$)} and even-in-$\omega$ {(symmetric with respect to $\omega$)} parts of the corresponding Green's functions which are in turn directly related to odd-$\omega$ and even-$\omega$ pairing amplitudes respectively. 
Note that, the quantity in Eq.~\eqref{anti-symmetry_in_I} can also be expressed as $\mathcal{B}^{2,1}_{\downarrow,\uparrow}=[\kappa_{\downarrow\uparrow}^{21}(\omega)-\kappa_{\downarrow\uparrow}^{21}(-\omega)]$. The voltage configuration necessary for measuring $\mathcal{A}$ and $\mathcal{B}$ is discussed above Eq.~\eqref{Eq2}. 
{It is worth mentioning that the Green's functions discussed above are not, in general, the free Green's function of the JJ, i.e. the Green's functions in the absence of tunneling probes. However, in the weak tunneling limit ($t_{1},t_{2}\ll1$), different inter-lead transmission amplitudes can be written in terms of free Green's functions (for details see SM Section C). Next, we will proceed to put the above qualitative discussion on firm grounds by calculating the quantity $\mathcal{A}^{2,1}_{\downarrow,\uparrow}$ from numerical analysis using a 2D model, and comparing the theoretical results discussed above (and also results presented in the SM). 
}

\begin{figure}[t]
	\centering{\includegraphics[width=0.45\textwidth]{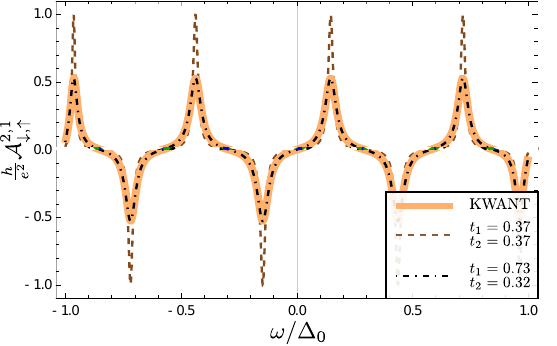}}
	\caption{\small \sl Anti-symmetry in non-local differential conductance $\mathcal{A}^{2,1}_{\downarrow,\uparrow}$  as a function of {$\omega=eV$, where $V$ is the applied bias voltage. Parameters are: $\mu=0$, $\phi_{21}=\pi/2$, $L=4.35\xi$, $\theta_1=0$, $\theta_2=\pi$.}}
	\label{Final_plot_KWANT}
\end{figure}

{\textit{Numerical simulation using 2D model:} In this part, we conduct a numerical simulation of a 2D lattice model depicted in Fig. \ref{device} using the KWANT package \cite{groth2014kwant} to further analyse the effective 1D model discussed above. The system is mainly a QSH insulator which is described by a real space configuration of the Bernevig-Hughes-Zhang (BHZ) model\cite{BHZ_SC_314_1757} mapped on a square lattice of dimension $140a\times140a$ with additional terms incorporated as needed. Here $a(=3\,nm$) is the lattice spacing. Two superconducting leads are attached to the bottom side ($S_1$) and to the right side ($S_2$) of the system, which are defined by the $s$-wave superconductivity-proximitized QSH insulators\cite{alicea_RPP_75_076501} having a finite phase difference ($\phi_{21}=\phi_2-\phi_1$) between them and thus, defining a 1D JJ extended over the left and top edges of the QSH region (see Fig.~\ref{device}(a)). On the other hand, two spin-polarized normal leads, which can be described by two quantum anomalous Hall insulators carrying chiral edge states of opposite chiralities, are tunnel-coupled via tiny insulating sections to the left side ($P_1$: having the $\uparrow$-spin channel only) and to the top side ($P_2$: having only the $\downarrow$-spin channel) of the QSH region. The QAH state of a spin-polarized lead is obtained by applying an exchange field to the QSH system, which oppositely affects the effective bulk gaps for the two spin states, eventually, at sufficient strength, destroying the topological state for one of them \cite{li_PRL_110_266802}. A strong onsite ferromagnetic impurity term is added over a small section of dimension $14a\times 14a$ at the extreme bottom-right corner (denoted by $M_x$) of the QSH region to ensure that the current is allowed to propagate only through the left and top edges of the QSH region in-between $S_1$ and $S_2$. A detailed insight into the 2D simulation is closely comparable to the same discussed in Ref.~\cite{adak_PhysicaE_139_115125} along with a comprehensive understanding of how particle-hole symmetry can be incorporated into the lattice model \cite{adak_PRB_106_045422}. For the numerical calculation, we consider the value of chemical potential ($\mu$) to be zero and the $s$-wave superconducting pairing potential ($\Delta_0$) as 2.0\,meV. The values of the remaining parameters related to the BHZ model closely resemble those describing HgTe/CdTe quantum wells\cite{BHZ_SC_314_1757,adak_PhysicaE_139_115125,adak_PRB_106_045422}. Finally, by conducting a numerical simulation with this setup, we compute the quantity $\mathcal{A}^{2,1}_{\downarrow,\uparrow}$ for $\phi_{21}=\pi/2$ as shown in FIG. \ref{Final_plot_KWANT} (solid cream line). The insulating necks of $P_1$ and $P_2$ are considered to be identical.}

{To compare the numerical result with the theoretical result, one needs to estimate the values of $L$, $t_1$, and $t_2$. To estimate the value of $L$, we analyzed the positions of the peaks and dips of $\mathcal{A}^{2,1}_{\downarrow,\uparrow}$. By mapping the corresponding energy values to the ABS energies, we estimate $L=4.35 \xi$ where $\xi=\hbar v_F/\Delta_0$ is the superconducting coherence length. Plugging this value of $L$ into theoretical calculations we numerically calculated the following quantity:
\begin{equation}
	\delta=\dfrac{1}{N}\sqrt{\sum_{n=1}^{N}\left(\mathcal{A}^{2,1}_{\downarrow,\uparrow}(n)|_{\text{Analytical}}-\mathcal{A}^{2,1}_{\downarrow,\uparrow}(n)|_{\text{Numerical}}\right)^2},
\end{equation}
for different values of $t_1$ and $t_2$. Here $N$ is the number of numerical grids within the energy window $-\Delta_0 \leq \omega \leq \Delta_0$. Keeping the tunneling strengths same, it turns out $\delta_{\text{min}} \approx 2\times 10^{-3}$ for $t_1=t_2=0.37$ (dashed brown line in FIG. \ref{Final_plot_KWANT}). Although, this plot is a good fit to the numerical results, if we relax the condition of $t_1=t_2$. then the value of $\delta$ can be even minimized giving rise to a better fit to the numerical results. It turns out, $\delta_{\text{min}} \approx 1\times 10^{-4}$ for $t_1=0.73$ and $t_2=0.32$ (black dot-dashed line in FIG. \ref{Final_plot_KWANT}). Dependence of $\mathcal{A}$ on the system parameters $t_1$, $t_2$ and $L$ is shown in SM Section D.}

{\textit{Discussion:} The above results rely on the fact that the spin-polarization axes of $P_1$ and $P_2$ can be adjusted exactly parallel or anti-parallel to the spin-polarization axis of HES.} For any other values of $\theta_1$ and $\theta_2$, $\mathcal{A}$ will have contributions from interference effects between the spin channels and the electron-hole channels due to the presence of more than one Green's function in the expressions of $t^{ij}_{p\bar{p}}$. These interference contributions will lead to the deviation from perfectly anti-symmetric features. Also, the detection of odd-frequency pairing through $\mathcal{A}$ does not require the probes $P_1$ and $P_2$ to be placed at different positions. This is due to the fact that the information of position enters into the calculation as pure phase contributions to $t^{ij}_{p\bar{p},s\bar{s}}$ ($\{s,\bar{s}\}\in \{ \uparrow,\downarrow \}$, $s\neq \bar{s}$).

{In our proposed set-up, the size of the sample (i.e. the length of the junction $L$) should be large enough to accommodate two spin-polarized probes. In Ref.~\cite{molenkamp_NP_8_485}, the width of the spin-polarized probe tip is $0.77$ $\mu m$, which sets the minimum dimension of the quantum spin-Hall sample (length of the edge) to be $1.54$ $\mu m$. However, in the same experiment, the size of the sample (length of the edge) is $4.1$ $\mu m$. Thus, it should be possible to experimentally fabricate a similar sample to realize our proposal. The values of other model parameters, e.g. Fermi velocity $v_F$, the chemical potential $\mu$, and pairing term $\Delta$ can be estimated from the experiment\cite{yacoby_NP_10_638}.}

Lastly, note that $\kappa_{he,\downarrow\uparrow}^{21}$ is related
to the ABS corresponding to the shuttling of a Cooper pair from left to right, while $\kappa_{he,\uparrow\downarrow}^{12}$ is related to the ABS
corresponding to the shuttling of Cooper pair in the opposite direction. Thus, in the weak tunneling limit, at $\varphi_{21}=0$ or $\pi$, (when these two ABS become degenerate), the differential conductances $\kappa_{he,\downarrow\uparrow}^{21}$ and $\kappa_{he,\uparrow\downarrow}^{12}$ becomes equal, leading to the vanishing of $\mathcal{A}^{2,1}_{\downarrow,\uparrow}$.

\textit{Acknowledgements:}
S.P. and A.M. acknowledge the Ministry of Education, India, and IISER Kolkata for funding. {V.A. would like to acknowledge Korea NRF (SRC Center for Quantum Coherence in Condensed Matter, Grant No. RS-2023-00207732) for funding.}

\textit{Author contribution:} S.P. and A.M. contributed equally to this work.

\bibliography{paper.bib}

\begin{thebibliography}{81}%
\makeatletter
\providecommand \@ifxundefined [1]{%
 \@ifx{#1\undefined}
}%
\providecommand \@ifnum [1]{%
 \ifnum #1\expandafter \@firstoftwo
 \else \expandafter \@secondoftwo
 \fi
}%
\providecommand \@ifx [1]{%
 \ifx #1\expandafter \@firstoftwo
 \else \expandafter \@secondoftwo
 \fi
}%
\providecommand \natexlab [1]{#1}%
\providecommand \enquote  [1]{``#1''}%
\providecommand \bibnamefont  [1]{#1}%
\providecommand \bibfnamefont [1]{#1}%
\providecommand \citenamefont [1]{#1}%
\providecommand \href@noop [0]{\@secondoftwo}%
\providecommand \href [0]{\begingroup \@sanitize@url \@href}%
\providecommand \@href[1]{\@@startlink{#1}\@@href}%
\providecommand \@@href[1]{\endgroup#1\@@endlink}%
\providecommand \@sanitize@url [0]{\catcode `\\12\catcode `\$12\catcode
  `\&12\catcode `\#12\catcode `\^12\catcode `\_12\catcode `\%12\relax}%
\providecommand \@@startlink[1]{}%
\providecommand \@@endlink[0]{}%
\providecommand \url  [0]{\begingroup\@sanitize@url \@url }%
\providecommand \@url [1]{\endgroup\@href {#1}{\urlprefix }}%
\providecommand \urlprefix  [0]{URL }%
\providecommand \Eprint [0]{\href }%
\providecommand \doibase [0]{https://doi.org/}%
\providecommand \selectlanguage [0]{\@gobble}%
\providecommand \bibinfo  [0]{\@secondoftwo}%
\providecommand \bibfield  [0]{\@secondoftwo}%
\providecommand \translation [1]{[#1]}%
\providecommand \BibitemOpen [0]{}%
\providecommand \bibitemStop [0]{}%
\providecommand \bibitemNoStop [0]{.\EOS\space}%
\providecommand \EOS [0]{\spacefactor3000\relax}%
\providecommand \BibitemShut  [1]{\csname bibitem#1\endcsname}%
\let\auto@bib@innerbib\@empty
\bibitem [{\citenamefont {Bardeen}\ \emph {et~al.}(1957)\citenamefont
  {Bardeen}, \citenamefont {Cooper},\ and\ \citenamefont {Schrieffer}}]{bard}%
  \BibitemOpen
  \bibfield  {author} {\bibinfo {author} {\bibfnamefont {J.}~\bibnamefont
  {Bardeen}}, \bibinfo {author} {\bibfnamefont {L.~N.}\ \bibnamefont
  {Cooper}},\ and\ \bibinfo {author} {\bibfnamefont {J.~R.}\ \bibnamefont
  {Schrieffer}},\ }\bibfield  {title} {\bibinfo {title} {Theory of
  superconductivity},\ }\href {https://doi.org/10.1103/PhysRev.108.1175}
  {\bibfield  {journal} {\bibinfo  {journal} {Phys. Rev.}\ }\textbf {\bibinfo
  {volume} {108}},\ \bibinfo {pages} {1175} (\bibinfo {year}
  {1957})}\BibitemShut {NoStop}%
\bibitem [{\citenamefont {Berezinskii}(1974)}]{bere}%
  \BibitemOpen
  \bibfield  {author} {\bibinfo {author} {\bibfnamefont {V.}~\bibnamefont
  {Berezinskii}},\ }\bibfield  {title} {\bibinfo {title} {New model of the
  anisotropic phase of superfluid he3},\ }\href@noop {} {\bibfield  {journal}
  {\bibinfo  {journal} {JETP Lett}\ }\textbf {\bibinfo {volume} {20}} (\bibinfo
  {year} {1974})}\BibitemShut {NoStop}%
\bibitem [{\citenamefont {Sigrist}\ and\ \citenamefont {Ueda}(1991)}]{sig}%
  \BibitemOpen
  \bibfield  {author} {\bibinfo {author} {\bibfnamefont {M.}~\bibnamefont
  {Sigrist}}\ and\ \bibinfo {author} {\bibfnamefont {K.}~\bibnamefont {Ueda}},\
  }\bibfield  {title} {\bibinfo {title} {Phenomenological theory of
  unconventional superconductivity},\ }\href
  {https://doi.org/10.1103/RevModPhys.63.239} {\bibfield  {journal} {\bibinfo
  {journal} {Rev. Mod. Phys.}\ }\textbf {\bibinfo {volume} {63}},\ \bibinfo
  {pages} {239} (\bibinfo {year} {1991})}\BibitemShut {NoStop}%
\bibitem [{\citenamefont {Linder}\ and\ \citenamefont {Balatsky}(2019)}]{lin}%
  \BibitemOpen
  \bibfield  {author} {\bibinfo {author} {\bibfnamefont {J.}~\bibnamefont
  {Linder}}\ and\ \bibinfo {author} {\bibfnamefont {A.~V.}\ \bibnamefont
  {Balatsky}},\ }\bibfield  {title} {\bibinfo {title} {Odd-frequency
  superconductivity},\ }\href {https://doi.org/10.1103/RevModPhys.91.045005}
  {\bibfield  {journal} {\bibinfo  {journal} {Rev. Mod. Phys.}\ }\textbf
  {\bibinfo {volume} {91}},\ \bibinfo {pages} {045005} (\bibinfo {year}
  {2019})}\BibitemShut {NoStop}%
\bibitem [{\citenamefont {Tanaka}\ \emph {et~al.}(2012)\citenamefont {Tanaka},
  \citenamefont {Sato},\ and\ \citenamefont {Nagaosa}}]{tana}%
  \BibitemOpen
  \bibfield  {author} {\bibinfo {author} {\bibfnamefont {Y.}~\bibnamefont
  {Tanaka}}, \bibinfo {author} {\bibfnamefont {M.}~\bibnamefont {Sato}},\ and\
  \bibinfo {author} {\bibfnamefont {N.}~\bibnamefont {Nagaosa}},\ }\bibfield
  {title} {\bibinfo {title} {Symmetry and topology in
  superconductors--odd-frequency pairing and edge states--},\ }\href@noop {}
  {\bibfield  {journal} {\bibinfo  {journal} {Journal of the Physical Society
  of Japan}\ }\textbf {\bibinfo {volume} {81}},\ \bibinfo {pages} {011013}
  (\bibinfo {year} {2012})}\BibitemShut {NoStop}%
\bibitem [{\citenamefont {Cayao}\ \emph {et~al.}(2020)\citenamefont {Cayao},
  \citenamefont {Triola},\ and\ \citenamefont {Black-Schaffer}}]{cay}%
  \BibitemOpen
  \bibfield  {author} {\bibinfo {author} {\bibfnamefont {J.}~\bibnamefont
  {Cayao}}, \bibinfo {author} {\bibfnamefont {C.}~\bibnamefont {Triola}},\ and\
  \bibinfo {author} {\bibfnamefont {A.~M.}\ \bibnamefont {Black-Schaffer}},\
  }\bibfield  {title} {\bibinfo {title} {Odd-frequency superconducting pairing
  in one-dimensional systems},\ }\href@noop {} {\bibfield  {journal} {\bibinfo
  {journal} {The European Physical Journal Special Topics}\ }\textbf {\bibinfo
  {volume} {229}},\ \bibinfo {pages} {545} (\bibinfo {year}
  {2020})}\BibitemShut {NoStop}%
\bibitem [{\citenamefont {Kirkpatrick}\ and\ \citenamefont
  {Belitz}(1991)}]{trk}%
  \BibitemOpen
  \bibfield  {author} {\bibinfo {author} {\bibfnamefont {T.~R.}\ \bibnamefont
  {Kirkpatrick}}\ and\ \bibinfo {author} {\bibfnamefont {D.}~\bibnamefont
  {Belitz}},\ }\bibfield  {title} {\bibinfo {title} {Disorder-induced triplet
  superconductivity},\ }\href {https://doi.org/10.1103/PhysRevLett.66.1533}
  {\bibfield  {journal} {\bibinfo  {journal} {Phys. Rev. Lett.}\ }\textbf
  {\bibinfo {volume} {66}},\ \bibinfo {pages} {1533} (\bibinfo {year}
  {1991})}\BibitemShut {NoStop}%
\bibitem [{\citenamefont {Belitz}\ and\ \citenamefont
  {Kirkpatrick}(1992)}]{bel}%
  \BibitemOpen
  \bibfield  {author} {\bibinfo {author} {\bibfnamefont {D.}~\bibnamefont
  {Belitz}}\ and\ \bibinfo {author} {\bibfnamefont {T.~R.}\ \bibnamefont
  {Kirkpatrick}},\ }\bibfield  {title} {\bibinfo {title} {Even-parity
  spin-triplet superconductivity in disordered electronic systems},\ }\href
  {https://doi.org/10.1103/PhysRevB.46.8393} {\bibfield  {journal} {\bibinfo
  {journal} {Phys. Rev. B}\ }\textbf {\bibinfo {volume} {46}},\ \bibinfo
  {pages} {8393} (\bibinfo {year} {1992})}\BibitemShut {NoStop}%
\bibitem [{\citenamefont {Balatsky}\ and\ \citenamefont
  {Abrahams}(1992)}]{bala}%
  \BibitemOpen
  \bibfield  {author} {\bibinfo {author} {\bibfnamefont {A.}~\bibnamefont
  {Balatsky}}\ and\ \bibinfo {author} {\bibfnamefont {E.}~\bibnamefont
  {Abrahams}},\ }\bibfield  {title} {\bibinfo {title} {New class of singlet
  superconductors which break the time reversal and parity},\ }\href
  {https://doi.org/10.1103/PhysRevB.45.13125} {\bibfield  {journal} {\bibinfo
  {journal} {Phys. Rev. B}\ }\textbf {\bibinfo {volume} {45}},\ \bibinfo
  {pages} {13125} (\bibinfo {year} {1992})}\BibitemShut {NoStop}%
\bibitem [{\citenamefont {Emery}\ and\ \citenamefont {Kivelson}(1992)}]{eme}%
  \BibitemOpen
  \bibfield  {author} {\bibinfo {author} {\bibfnamefont {V.~J.}\ \bibnamefont
  {Emery}}\ and\ \bibinfo {author} {\bibfnamefont {S.}~\bibnamefont
  {Kivelson}},\ }\bibfield  {title} {\bibinfo {title} {Mapping of the
  two-channel kondo problem to a resonant-level model},\ }\href
  {https://doi.org/10.1103/PhysRevB.46.10812} {\bibfield  {journal} {\bibinfo
  {journal} {Phys. Rev. B}\ }\textbf {\bibinfo {volume} {46}},\ \bibinfo
  {pages} {10812} (\bibinfo {year} {1992})}\BibitemShut {NoStop}%
\bibitem [{\citenamefont {Balatsky}\ and\ \citenamefont
  {Bonc\ifmmode~\check{}\else \v{}\fi{}a}(1993)}]{bon}%
  \BibitemOpen
  \bibfield  {author} {\bibinfo {author} {\bibfnamefont {A.~V.}\ \bibnamefont
  {Balatsky}}\ and\ \bibinfo {author} {\bibfnamefont {J.}~\bibnamefont
  {Bonc\ifmmode~\check{}\else \v{}\fi{}a}},\ }\bibfield  {title} {\bibinfo
  {title} {Even- and odd-frequency pairing correlations in the one-dimensional
  t-j-h model: A comparative study},\ }\href
  {https://doi.org/10.1103/PhysRevB.48.7445} {\bibfield  {journal} {\bibinfo
  {journal} {Phys. Rev. B}\ }\textbf {\bibinfo {volume} {48}},\ \bibinfo
  {pages} {7445} (\bibinfo {year} {1993})}\BibitemShut {NoStop}%
\bibitem [{\citenamefont {Bulut}\ \emph {et~al.}(1993)\citenamefont {Bulut},
  \citenamefont {Scalapino},\ and\ \citenamefont {White}}]{bulu}%
  \BibitemOpen
  \bibfield  {author} {\bibinfo {author} {\bibfnamefont {N.}~\bibnamefont
  {Bulut}}, \bibinfo {author} {\bibfnamefont {D.~J.}\ \bibnamefont
  {Scalapino}},\ and\ \bibinfo {author} {\bibfnamefont {S.~R.}\ \bibnamefont
  {White}},\ }\bibfield  {title} {\bibinfo {title} {Effective particle-particle
  interaction in the two-dimensional hubbard model},\ }\href
  {https://doi.org/10.1103/PhysRevB.47.6157} {\bibfield  {journal} {\bibinfo
  {journal} {Phys. Rev. B}\ }\textbf {\bibinfo {volume} {47}},\ \bibinfo
  {pages} {6157} (\bibinfo {year} {1993})}\BibitemShut {NoStop}%
\bibitem [{\citenamefont {Coleman}\ \emph {et~al.}(1993)\citenamefont
  {Coleman}, \citenamefont {Miranda},\ and\ \citenamefont {Tsvelik}}]{cole}%
  \BibitemOpen
  \bibfield  {author} {\bibinfo {author} {\bibfnamefont {P.}~\bibnamefont
  {Coleman}}, \bibinfo {author} {\bibfnamefont {E.}~\bibnamefont {Miranda}},\
  and\ \bibinfo {author} {\bibfnamefont {A.}~\bibnamefont {Tsvelik}},\
  }\bibfield  {title} {\bibinfo {title} {Possible realization of odd-frequency
  pairing in heavy fermion compounds},\ }\href
  {https://doi.org/10.1103/PhysRevLett.70.2960} {\bibfield  {journal} {\bibinfo
   {journal} {Phys. Rev. Lett.}\ }\textbf {\bibinfo {volume} {70}},\ \bibinfo
  {pages} {2960} (\bibinfo {year} {1993})}\BibitemShut {NoStop}%
\bibitem [{\citenamefont {Kashuba}\ \emph {et~al.}(2017)\citenamefont
  {Kashuba}, \citenamefont {Sothmann}, \citenamefont {Burset},\ and\
  \citenamefont {Trauzettel}}]{oka}%
  \BibitemOpen
  \bibfield  {author} {\bibinfo {author} {\bibfnamefont {O.}~\bibnamefont
  {Kashuba}}, \bibinfo {author} {\bibfnamefont {B.}~\bibnamefont {Sothmann}},
  \bibinfo {author} {\bibfnamefont {P.}~\bibnamefont {Burset}},\ and\ \bibinfo
  {author} {\bibfnamefont {B.}~\bibnamefont {Trauzettel}},\ }\bibfield  {title}
  {\bibinfo {title} {Majorana stm as a perfect detector of odd-frequency
  superconductivity},\ }\href {https://doi.org/10.1103/PhysRevB.95.174516}
  {\bibfield  {journal} {\bibinfo  {journal} {Phys. Rev. B}\ }\textbf {\bibinfo
  {volume} {95}},\ \bibinfo {pages} {174516} (\bibinfo {year}
  {2017})}\BibitemShut {NoStop}%
\bibitem [{\citenamefont {Schemm}\ \emph {et~al.}(2014)\citenamefont {Schemm},
  \citenamefont {Gannon}, \citenamefont {Wishne}, \citenamefont {Halperin},\
  and\ \citenamefont {Kapitulnik}}]{ers}%
  \BibitemOpen
  \bibfield  {author} {\bibinfo {author} {\bibfnamefont {E.}~\bibnamefont
  {Schemm}}, \bibinfo {author} {\bibfnamefont {W.}~\bibnamefont {Gannon}},
  \bibinfo {author} {\bibfnamefont {C.}~\bibnamefont {Wishne}}, \bibinfo
  {author} {\bibfnamefont {W.}~\bibnamefont {Halperin}},\ and\ \bibinfo
  {author} {\bibfnamefont {A.}~\bibnamefont {Kapitulnik}},\ }\bibfield  {title}
  {\bibinfo {title} {Observation of broken time-reversal symmetry in the
  heavy-fermion superconductor upt3},\ }\href@noop {} {\bibfield  {journal}
  {\bibinfo  {journal} {Science}\ }\textbf {\bibinfo {volume} {345}},\ \bibinfo
  {pages} {190} (\bibinfo {year} {2014})}\BibitemShut {NoStop}%
\bibitem [{\citenamefont {Komendov\'a}\ and\ \citenamefont
  {Black-Schaffer}(2017)}]{kom}%
  \BibitemOpen
  \bibfield  {author} {\bibinfo {author} {\bibfnamefont {L.}~\bibnamefont
  {Komendov\'a}}\ and\ \bibinfo {author} {\bibfnamefont {A.~M.}\ \bibnamefont
  {Black-Schaffer}},\ }\bibfield  {title} {\bibinfo {title} {Odd-frequency
  superconductivity in ${\mathrm{sr}}_{2}{\mathrm{ruo}}_{4}$ measured by kerr
  rotation},\ }\href {https://doi.org/10.1103/PhysRevLett.119.087001}
  {\bibfield  {journal} {\bibinfo  {journal} {Phys. Rev. Lett.}\ }\textbf
  {\bibinfo {volume} {119}},\ \bibinfo {pages} {087001} (\bibinfo {year}
  {2017})}\BibitemShut {NoStop}%
\bibitem [{\citenamefont {Di~Bernardo}\ \emph
  {et~al.}(2015{\natexlab{a}})\citenamefont {Di~Bernardo}, \citenamefont
  {Salman}, \citenamefont {Wang}, \citenamefont {Amado}, \citenamefont
  {Egilmez}, \citenamefont {Flokstra}, \citenamefont {Suter}, \citenamefont
  {Lee}, \citenamefont {Zhao}, \citenamefont {Prokscha}, \citenamefont
  {Morenzoni}, \citenamefont {Blamire}, \citenamefont {Linder},\ and\
  \citenamefont {Robinson}}]{adi}%
  \BibitemOpen
  \bibfield  {author} {\bibinfo {author} {\bibfnamefont {A.}~\bibnamefont
  {Di~Bernardo}}, \bibinfo {author} {\bibfnamefont {Z.}~\bibnamefont {Salman}},
  \bibinfo {author} {\bibfnamefont {X.~L.}\ \bibnamefont {Wang}}, \bibinfo
  {author} {\bibfnamefont {M.}~\bibnamefont {Amado}}, \bibinfo {author}
  {\bibfnamefont {M.}~\bibnamefont {Egilmez}}, \bibinfo {author} {\bibfnamefont
  {M.~G.}\ \bibnamefont {Flokstra}}, \bibinfo {author} {\bibfnamefont
  {A.}~\bibnamefont {Suter}}, \bibinfo {author} {\bibfnamefont {S.~L.}\
  \bibnamefont {Lee}}, \bibinfo {author} {\bibfnamefont {J.~H.}\ \bibnamefont
  {Zhao}}, \bibinfo {author} {\bibfnamefont {T.}~\bibnamefont {Prokscha}},
  \bibinfo {author} {\bibfnamefont {E.}~\bibnamefont {Morenzoni}}, \bibinfo
  {author} {\bibfnamefont {M.~G.}\ \bibnamefont {Blamire}}, \bibinfo {author}
  {\bibfnamefont {J.}~\bibnamefont {Linder}},\ and\ \bibinfo {author}
  {\bibfnamefont {J.~W.~A.}\ \bibnamefont {Robinson}},\ }\bibfield  {title}
  {\bibinfo {title} {Intrinsic paramagnetic meissner effect due to $s$-wave
  odd-frequency superconductivity},\ }\href
  {https://doi.org/10.1103/PhysRevX.5.041021} {\bibfield  {journal} {\bibinfo
  {journal} {Phys. Rev. X}\ }\textbf {\bibinfo {volume} {5}},\ \bibinfo {pages}
  {041021} (\bibinfo {year} {2015}{\natexlab{a}})}\BibitemShut {NoStop}%
\bibitem [{\citenamefont {Bergeret}\ \emph
  {et~al.}(2001{\natexlab{a}})\citenamefont {Bergeret}, \citenamefont
  {Volkov},\ and\ \citenamefont {Efetov}}]{bse}%
  \BibitemOpen
  \bibfield  {author} {\bibinfo {author} {\bibfnamefont {F.~S.}\ \bibnamefont
  {Bergeret}}, \bibinfo {author} {\bibfnamefont {A.~F.}\ \bibnamefont
  {Volkov}},\ and\ \bibinfo {author} {\bibfnamefont {K.~B.}\ \bibnamefont
  {Efetov}},\ }\bibfield  {title} {\bibinfo {title} {Josephson current in
  superconductor-ferromagnet structures with a nonhomogeneous magnetization},\
  }\href {https://doi.org/10.1103/PhysRevB.64.134506} {\bibfield  {journal}
  {\bibinfo  {journal} {Phys. Rev. B}\ }\textbf {\bibinfo {volume} {64}},\
  \bibinfo {pages} {134506} (\bibinfo {year} {2001}{\natexlab{a}})}\BibitemShut
  {NoStop}%
\bibitem [{\citenamefont {Alidoust}\ \emph {et~al.}(2014)\citenamefont
  {Alidoust}, \citenamefont {Halterman},\ and\ \citenamefont {Linder}}]{ali}%
  \BibitemOpen
  \bibfield  {author} {\bibinfo {author} {\bibfnamefont {M.}~\bibnamefont
  {Alidoust}}, \bibinfo {author} {\bibfnamefont {K.}~\bibnamefont
  {Halterman}},\ and\ \bibinfo {author} {\bibfnamefont {J.}~\bibnamefont
  {Linder}},\ }\bibfield  {title} {\bibinfo {title} {Meissner effect probing of
  odd-frequency triplet pairing in superconducting spin valves},\ }\href
  {https://doi.org/10.1103/PhysRevB.89.054508} {\bibfield  {journal} {\bibinfo
  {journal} {Phys. Rev. B}\ }\textbf {\bibinfo {volume} {89}},\ \bibinfo
  {pages} {054508} (\bibinfo {year} {2014})}\BibitemShut {NoStop}%
\bibitem [{\citenamefont {Krieger}\ \emph {et~al.}(2020)\citenamefont
  {Krieger}, \citenamefont {Pertsova}, \citenamefont {Giblin}, \citenamefont
  {D\"obeli}, \citenamefont {Prokscha}, \citenamefont {Schneider},
  \citenamefont {Suter}, \citenamefont {Hesjedal}, \citenamefont {Balatsky},\
  and\ \citenamefont {Salman}}]{jak}%
  \BibitemOpen
  \bibfield  {author} {\bibinfo {author} {\bibfnamefont {J.~A.}\ \bibnamefont
  {Krieger}}, \bibinfo {author} {\bibfnamefont {A.}~\bibnamefont {Pertsova}},
  \bibinfo {author} {\bibfnamefont {S.~R.}\ \bibnamefont {Giblin}}, \bibinfo
  {author} {\bibfnamefont {M.}~\bibnamefont {D\"obeli}}, \bibinfo {author}
  {\bibfnamefont {T.}~\bibnamefont {Prokscha}}, \bibinfo {author}
  {\bibfnamefont {C.~W.}\ \bibnamefont {Schneider}}, \bibinfo {author}
  {\bibfnamefont {A.}~\bibnamefont {Suter}}, \bibinfo {author} {\bibfnamefont
  {T.}~\bibnamefont {Hesjedal}}, \bibinfo {author} {\bibfnamefont {A.~V.}\
  \bibnamefont {Balatsky}},\ and\ \bibinfo {author} {\bibfnamefont
  {Z.}~\bibnamefont {Salman}},\ }\bibfield  {title} {\bibinfo {title}
  {Proximity-induced odd-frequency superconductivity in a topological
  insulator},\ }\href {https://doi.org/10.1103/PhysRevLett.125.026802}
  {\bibfield  {journal} {\bibinfo  {journal} {Phys. Rev. Lett.}\ }\textbf
  {\bibinfo {volume} {125}},\ \bibinfo {pages} {026802} (\bibinfo {year}
  {2020})}\BibitemShut {NoStop}%
\bibitem [{\citenamefont {Yokoyama}\ \emph {et~al.}(2011)\citenamefont
  {Yokoyama}, \citenamefont {Tanaka},\ and\ \citenamefont {Nagaosa}}]{yta}%
  \BibitemOpen
  \bibfield  {author} {\bibinfo {author} {\bibfnamefont {T.}~\bibnamefont
  {Yokoyama}}, \bibinfo {author} {\bibfnamefont {Y.}~\bibnamefont {Tanaka}},\
  and\ \bibinfo {author} {\bibfnamefont {N.}~\bibnamefont {Nagaosa}},\
  }\bibfield  {title} {\bibinfo {title} {Anomalous meissner effect in a
  normal-metal--superconductor junction with a spin-active interface},\ }\href
  {https://doi.org/10.1103/PhysRevLett.106.246601} {\bibfield  {journal}
  {\bibinfo  {journal} {Phys. Rev. Lett.}\ }\textbf {\bibinfo {volume} {106}},\
  \bibinfo {pages} {246601} (\bibinfo {year} {2011})}\BibitemShut {NoStop}%
\bibitem [{\citenamefont {Perrin}\ \emph {et~al.}(2020)\citenamefont {Perrin},
  \citenamefont {Santos}, \citenamefont {M\'enard}, \citenamefont {Brun},
  \citenamefont {Cren}, \citenamefont {Civelli},\ and\ \citenamefont
  {Simon}}]{perr}%
  \BibitemOpen
  \bibfield  {author} {\bibinfo {author} {\bibfnamefont {V.}~\bibnamefont
  {Perrin}}, \bibinfo {author} {\bibfnamefont {F.~L.~N.}\ \bibnamefont
  {Santos}}, \bibinfo {author} {\bibfnamefont {G.~C.}\ \bibnamefont
  {M\'enard}}, \bibinfo {author} {\bibfnamefont {C.}~\bibnamefont {Brun}},
  \bibinfo {author} {\bibfnamefont {T.}~\bibnamefont {Cren}}, \bibinfo {author}
  {\bibfnamefont {M.}~\bibnamefont {Civelli}},\ and\ \bibinfo {author}
  {\bibfnamefont {P.}~\bibnamefont {Simon}},\ }\bibfield  {title} {\bibinfo
  {title} {Unveiling odd-frequency pairing around a magnetic impurity in a
  superconductor},\ }\href {https://doi.org/10.1103/PhysRevLett.125.117003}
  {\bibfield  {journal} {\bibinfo  {journal} {Phys. Rev. Lett.}\ }\textbf
  {\bibinfo {volume} {125}},\ \bibinfo {pages} {117003} (\bibinfo {year}
  {2020})}\BibitemShut {NoStop}%
\bibitem [{\citenamefont {Kuzmanovski}\ \emph {et~al.}(2020)\citenamefont
  {Kuzmanovski}, \citenamefont {Souto},\ and\ \citenamefont {Balatsky}}]{kuzz}%
  \BibitemOpen
  \bibfield  {author} {\bibinfo {author} {\bibfnamefont {D.}~\bibnamefont
  {Kuzmanovski}}, \bibinfo {author} {\bibfnamefont {R.~S.}\ \bibnamefont
  {Souto}},\ and\ \bibinfo {author} {\bibfnamefont {A.~V.}\ \bibnamefont
  {Balatsky}},\ }\bibfield  {title} {\bibinfo {title} {Odd-frequency
  superconductivity near a magnetic impurity in a conventional
  superconductor},\ }\href {https://doi.org/10.1103/PhysRevB.101.094505}
  {\bibfield  {journal} {\bibinfo  {journal} {Phys. Rev. B}\ }\textbf {\bibinfo
  {volume} {101}},\ \bibinfo {pages} {094505} (\bibinfo {year}
  {2020})}\BibitemShut {NoStop}%
\bibitem [{\citenamefont {Kornich}\ \emph {et~al.}(2021)\citenamefont
  {Kornich}, \citenamefont {Schlawin}, \citenamefont {Sentef},\ and\
  \citenamefont {Trauzettel}}]{vko}%
  \BibitemOpen
  \bibfield  {author} {\bibinfo {author} {\bibfnamefont {V.}~\bibnamefont
  {Kornich}}, \bibinfo {author} {\bibfnamefont {F.}~\bibnamefont {Schlawin}},
  \bibinfo {author} {\bibfnamefont {M.~A.}\ \bibnamefont {Sentef}},\ and\
  \bibinfo {author} {\bibfnamefont {B.}~\bibnamefont {Trauzettel}},\ }\bibfield
   {title} {\bibinfo {title} {Direct detection of odd-frequency
  superconductivity via time- and angle-resolved photoelectron fluctuation
  spectroscopy},\ }\href {https://doi.org/10.1103/PhysRevResearch.3.L042034}
  {\bibfield  {journal} {\bibinfo  {journal} {Phys. Rev. Res.}\ }\textbf
  {\bibinfo {volume} {3}},\ \bibinfo {pages} {L042034} (\bibinfo {year}
  {2021})}\BibitemShut {NoStop}%
\bibitem [{\citenamefont {Chakraborty}\ and\ \citenamefont
  {Black-Schaffer}(2022)}]{dca}%
  \BibitemOpen
  \bibfield  {author} {\bibinfo {author} {\bibfnamefont {D.}~\bibnamefont
  {Chakraborty}}\ and\ \bibinfo {author} {\bibfnamefont {A.~M.}\ \bibnamefont
  {Black-Schaffer}},\ }\bibfield  {title} {\bibinfo {title} {Quasiparticle
  interference as a direct experimental probe of bulk odd-frequency
  superconducting pairing},\ }\href
  {https://doi.org/10.1103/PhysRevLett.129.247001} {\bibfield  {journal}
  {\bibinfo  {journal} {Phys. Rev. Lett.}\ }\textbf {\bibinfo {volume} {129}},\
  \bibinfo {pages} {247001} (\bibinfo {year} {2022})}\BibitemShut {NoStop}%
\bibitem [{\citenamefont {Abrahams}\ \emph {et~al.}(1995)\citenamefont
  {Abrahams}, \citenamefont {Balatsky}, \citenamefont {Scalapino},\ and\
  \citenamefont {Schrieffer}}]{eab}%
  \BibitemOpen
  \bibfield  {author} {\bibinfo {author} {\bibfnamefont {E.}~\bibnamefont
  {Abrahams}}, \bibinfo {author} {\bibfnamefont {A.}~\bibnamefont {Balatsky}},
  \bibinfo {author} {\bibfnamefont {D.~J.}\ \bibnamefont {Scalapino}},\ and\
  \bibinfo {author} {\bibfnamefont {J.~R.}\ \bibnamefont {Schrieffer}},\
  }\bibfield  {title} {\bibinfo {title} {Properties of odd-gap
  superconductors},\ }\href {https://doi.org/10.1103/PhysRevB.52.1271}
  {\bibfield  {journal} {\bibinfo  {journal} {Phys. Rev. B}\ }\textbf {\bibinfo
  {volume} {52}},\ \bibinfo {pages} {1271} (\bibinfo {year}
  {1995})}\BibitemShut {NoStop}%
\bibitem [{\citenamefont {Bergeret}\ \emph
  {et~al.}(2001{\natexlab{b}})\citenamefont {Bergeret}, \citenamefont
  {Volkov},\ and\ \citenamefont {Efetov}}]{berg1}%
  \BibitemOpen
  \bibfield  {author} {\bibinfo {author} {\bibfnamefont {F.~S.}\ \bibnamefont
  {Bergeret}}, \bibinfo {author} {\bibfnamefont {A.~F.}\ \bibnamefont
  {Volkov}},\ and\ \bibinfo {author} {\bibfnamefont {K.~B.}\ \bibnamefont
  {Efetov}},\ }\bibfield  {title} {\bibinfo {title} {Long-range proximity
  effects in superconductor-ferromagnet structures},\ }\href
  {https://doi.org/10.1103/PhysRevLett.86.4096} {\bibfield  {journal} {\bibinfo
   {journal} {Phys. Rev. Lett.}\ }\textbf {\bibinfo {volume} {86}},\ \bibinfo
  {pages} {4096} (\bibinfo {year} {2001}{\natexlab{b}})}\BibitemShut {NoStop}%
\bibitem [{\citenamefont {Bergeret}\ \emph {et~al.}(2003)\citenamefont
  {Bergeret}, \citenamefont {Volkov},\ and\ \citenamefont {Efetov}}]{berg2}%
  \BibitemOpen
  \bibfield  {author} {\bibinfo {author} {\bibfnamefont {F.~S.}\ \bibnamefont
  {Bergeret}}, \bibinfo {author} {\bibfnamefont {A.~F.}\ \bibnamefont
  {Volkov}},\ and\ \bibinfo {author} {\bibfnamefont {K.~B.}\ \bibnamefont
  {Efetov}},\ }\bibfield  {title} {\bibinfo {title} {Manifestation of triplet
  superconductivity in superconductor-ferromagnet structures},\ }\href
  {https://doi.org/10.1103/PhysRevB.68.064513} {\bibfield  {journal} {\bibinfo
  {journal} {Phys. Rev. B}\ }\textbf {\bibinfo {volume} {68}},\ \bibinfo
  {pages} {064513} (\bibinfo {year} {2003})}\BibitemShut {NoStop}%
\bibitem [{\citenamefont {Bergeret}\ \emph {et~al.}(2005)\citenamefont
  {Bergeret}, \citenamefont {Volkov},\ and\ \citenamefont {Efetov}}]{berg3}%
  \BibitemOpen
  \bibfield  {author} {\bibinfo {author} {\bibfnamefont {F.~S.}\ \bibnamefont
  {Bergeret}}, \bibinfo {author} {\bibfnamefont {A.~F.}\ \bibnamefont
  {Volkov}},\ and\ \bibinfo {author} {\bibfnamefont {K.~B.}\ \bibnamefont
  {Efetov}},\ }\bibfield  {title} {\bibinfo {title} {Odd triplet
  superconductivity and related phenomena in superconductor-ferromagnet
  structures},\ }\href {https://doi.org/10.1103/RevModPhys.77.1321} {\bibfield
  {journal} {\bibinfo  {journal} {Rev. Mod. Phys.}\ }\textbf {\bibinfo {volume}
  {77}},\ \bibinfo {pages} {1321} (\bibinfo {year} {2005})}\BibitemShut
  {NoStop}%
\bibitem [{\citenamefont {Eschrig}\ \emph {et~al.}(2003)\citenamefont
  {Eschrig}, \citenamefont {Kopu}, \citenamefont {Cuevas},\ and\ \citenamefont
  {Sch\"on}}]{kopu}%
  \BibitemOpen
  \bibfield  {author} {\bibinfo {author} {\bibfnamefont {M.}~\bibnamefont
  {Eschrig}}, \bibinfo {author} {\bibfnamefont {J.}~\bibnamefont {Kopu}},
  \bibinfo {author} {\bibfnamefont {J.~C.}\ \bibnamefont {Cuevas}},\ and\
  \bibinfo {author} {\bibfnamefont {G.}~\bibnamefont {Sch\"on}},\ }\bibfield
  {title} {\bibinfo {title} {Theory of half-metal/superconductor
  heterostructures},\ }\href {https://doi.org/10.1103/PhysRevLett.90.137003}
  {\bibfield  {journal} {\bibinfo  {journal} {Phys. Rev. Lett.}\ }\textbf
  {\bibinfo {volume} {90}},\ \bibinfo {pages} {137003} (\bibinfo {year}
  {2003})}\BibitemShut {NoStop}%
\bibitem [{\citenamefont {Volkov}\ \emph {et~al.}(2006)\citenamefont {Volkov},
  \citenamefont {Anishchanka},\ and\ \citenamefont {Efetov}}]{vol}%
  \BibitemOpen
  \bibfield  {author} {\bibinfo {author} {\bibfnamefont {A.~F.}\ \bibnamefont
  {Volkov}}, \bibinfo {author} {\bibfnamefont {A.}~\bibnamefont
  {Anishchanka}},\ and\ \bibinfo {author} {\bibfnamefont {K.~B.}\ \bibnamefont
  {Efetov}},\ }\bibfield  {title} {\bibinfo {title} {Odd triplet
  superconductivity in a superconductor/ferromagnet system with a spiral
  magnetic structure},\ }\href {https://doi.org/10.1103/PhysRevB.73.104412}
  {\bibfield  {journal} {\bibinfo  {journal} {Phys. Rev. B}\ }\textbf {\bibinfo
  {volume} {73}},\ \bibinfo {pages} {104412} (\bibinfo {year}
  {2006})}\BibitemShut {NoStop}%
\bibitem [{\citenamefont {Yokoyama}\ \emph {et~al.}(2007)\citenamefont
  {Yokoyama}, \citenamefont {Tanaka},\ and\ \citenamefont {Golubov}}]{toko}%
  \BibitemOpen
  \bibfield  {author} {\bibinfo {author} {\bibfnamefont {T.}~\bibnamefont
  {Yokoyama}}, \bibinfo {author} {\bibfnamefont {Y.}~\bibnamefont {Tanaka}},\
  and\ \bibinfo {author} {\bibfnamefont {A.~A.}\ \bibnamefont {Golubov}},\
  }\bibfield  {title} {\bibinfo {title} {Manifestation of the odd-frequency
  spin-triplet pairing state in diffusive ferromagnet/superconductor
  junctions},\ }\href {https://doi.org/10.1103/PhysRevB.75.134510} {\bibfield
  {journal} {\bibinfo  {journal} {Phys. Rev. B}\ }\textbf {\bibinfo {volume}
  {75}},\ \bibinfo {pages} {134510} (\bibinfo {year} {2007})}\BibitemShut
  {NoStop}%
\bibitem [{\citenamefont {Tanaka}\ \emph
  {et~al.}(2007{\natexlab{a}})\citenamefont {Tanaka}, \citenamefont {Tanuma},\
  and\ \citenamefont {Golubov}}]{tan}%
  \BibitemOpen
  \bibfield  {author} {\bibinfo {author} {\bibfnamefont {Y.}~\bibnamefont
  {Tanaka}}, \bibinfo {author} {\bibfnamefont {Y.}~\bibnamefont {Tanuma}},\
  and\ \bibinfo {author} {\bibfnamefont {A.~A.}\ \bibnamefont {Golubov}},\
  }\bibfield  {title} {\bibinfo {title} {Odd-frequency pairing in
  normal-metal/superconductor junctions},\ }\href
  {https://doi.org/10.1103/PhysRevB.76.054522} {\bibfield  {journal} {\bibinfo
  {journal} {Phys. Rev. B}\ }\textbf {\bibinfo {volume} {76}},\ \bibinfo
  {pages} {054522} (\bibinfo {year} {2007}{\natexlab{a}})}\BibitemShut
  {NoStop}%
\bibitem [{\citenamefont {Black-Schaffer}\ and\ \citenamefont
  {Balatsky}(2012)}]{amb1}%
  \BibitemOpen
  \bibfield  {author} {\bibinfo {author} {\bibfnamefont {A.~M.}\ \bibnamefont
  {Black-Schaffer}}\ and\ \bibinfo {author} {\bibfnamefont {A.~V.}\
  \bibnamefont {Balatsky}},\ }\bibfield  {title} {\bibinfo {title}
  {Odd-frequency superconducting pairing in topological insulators},\ }\href
  {https://doi.org/10.1103/PhysRevB.86.144506} {\bibfield  {journal} {\bibinfo
  {journal} {Phys. Rev. B}\ }\textbf {\bibinfo {volume} {86}},\ \bibinfo
  {pages} {144506} (\bibinfo {year} {2012})}\BibitemShut {NoStop}%
\bibitem [{\citenamefont {Black-Schaffer}\ and\ \citenamefont
  {Balatsky}(2013)}]{amb2}%
  \BibitemOpen
  \bibfield  {author} {\bibinfo {author} {\bibfnamefont {A.~M.}\ \bibnamefont
  {Black-Schaffer}}\ and\ \bibinfo {author} {\bibfnamefont {A.~V.}\
  \bibnamefont {Balatsky}},\ }\bibfield  {title} {\bibinfo {title}
  {Proximity-induced unconventional superconductivity in topological
  insulators},\ }\href {https://doi.org/10.1103/PhysRevB.87.220506} {\bibfield
  {journal} {\bibinfo  {journal} {Phys. Rev. B}\ }\textbf {\bibinfo {volume}
  {87}},\ \bibinfo {pages} {220506} (\bibinfo {year} {2013})}\BibitemShut
  {NoStop}%
\bibitem [{\citenamefont {Cr\'epin}\ \emph {et~al.}(2015)\citenamefont
  {Cr\'epin}, \citenamefont {Burset},\ and\ \citenamefont {Trauzettel}}]{cre}%
  \BibitemOpen
  \bibfield  {author} {\bibinfo {author} {\bibfnamefont {F.~m.~c.}\
  \bibnamefont {Cr\'epin}}, \bibinfo {author} {\bibfnamefont {P.}~\bibnamefont
  {Burset}},\ and\ \bibinfo {author} {\bibfnamefont {B.}~\bibnamefont
  {Trauzettel}},\ }\bibfield  {title} {\bibinfo {title} {Odd-frequency triplet
  superconductivity at the helical edge of a topological insulator},\ }\href
  {https://doi.org/10.1103/PhysRevB.92.100507} {\bibfield  {journal} {\bibinfo
  {journal} {Phys. Rev. B}\ }\textbf {\bibinfo {volume} {92}},\ \bibinfo
  {pages} {100507} (\bibinfo {year} {2015})}\BibitemShut {NoStop}%
\bibitem [{\citenamefont {Burset}\ \emph {et~al.}(2015)\citenamefont {Burset},
  \citenamefont {Lu}, \citenamefont {Tkachov}, \citenamefont {Tanaka},
  \citenamefont {Hankiewicz},\ and\ \citenamefont {Trauzettel}}]{burr}%
  \BibitemOpen
  \bibfield  {author} {\bibinfo {author} {\bibfnamefont {P.}~\bibnamefont
  {Burset}}, \bibinfo {author} {\bibfnamefont {B.}~\bibnamefont {Lu}}, \bibinfo
  {author} {\bibfnamefont {G.}~\bibnamefont {Tkachov}}, \bibinfo {author}
  {\bibfnamefont {Y.}~\bibnamefont {Tanaka}}, \bibinfo {author} {\bibfnamefont
  {E.~M.}\ \bibnamefont {Hankiewicz}},\ and\ \bibinfo {author} {\bibfnamefont
  {B.}~\bibnamefont {Trauzettel}},\ }\bibfield  {title} {\bibinfo {title}
  {Superconducting proximity effect in three-dimensional topological insulators
  in the presence of a magnetic field},\ }\href
  {https://doi.org/10.1103/PhysRevB.92.205424} {\bibfield  {journal} {\bibinfo
  {journal} {Phys. Rev. B}\ }\textbf {\bibinfo {volume} {92}},\ \bibinfo
  {pages} {205424} (\bibinfo {year} {2015})}\BibitemShut {NoStop}%
\bibitem [{\citenamefont {Cayao}\ and\ \citenamefont
  {Black-Schaffer}(2017)}]{cayao}%
  \BibitemOpen
  \bibfield  {author} {\bibinfo {author} {\bibfnamefont {J.}~\bibnamefont
  {Cayao}}\ and\ \bibinfo {author} {\bibfnamefont {A.~M.}\ \bibnamefont
  {Black-Schaffer}},\ }\bibfield  {title} {\bibinfo {title} {Odd-frequency
  superconducting pairing and subgap density of states at the edge of a
  two-dimensional topological insulator without magnetism},\ }\href
  {https://doi.org/10.1103/PhysRevB.96.155426} {\bibfield  {journal} {\bibinfo
  {journal} {Phys. Rev. B}\ }\textbf {\bibinfo {volume} {96}},\ \bibinfo
  {pages} {155426} (\bibinfo {year} {2017})}\BibitemShut {NoStop}%
\bibitem [{\citenamefont {Hwang}\ \emph {et~al.}(2018)\citenamefont {Hwang},
  \citenamefont {Burset},\ and\ \citenamefont {Sothmann}}]{hwa}%
  \BibitemOpen
  \bibfield  {author} {\bibinfo {author} {\bibfnamefont {S.-Y.}\ \bibnamefont
  {Hwang}}, \bibinfo {author} {\bibfnamefont {P.}~\bibnamefont {Burset}},\ and\
  \bibinfo {author} {\bibfnamefont {B.}~\bibnamefont {Sothmann}},\ }\bibfield
  {title} {\bibinfo {title} {Odd-frequency superconductivity revealed by
  thermopower},\ }\href {https://doi.org/10.1103/PhysRevB.98.161408} {\bibfield
   {journal} {\bibinfo  {journal} {Phys. Rev. B}\ }\textbf {\bibinfo {volume}
  {98}},\ \bibinfo {pages} {161408} (\bibinfo {year} {2018})}\BibitemShut
  {NoStop}%
\bibitem [{\citenamefont {Cayao}\ and\ \citenamefont
  {Black-Schaffer}(2018)}]{cay1}%
  \BibitemOpen
  \bibfield  {author} {\bibinfo {author} {\bibfnamefont {J.}~\bibnamefont
  {Cayao}}\ and\ \bibinfo {author} {\bibfnamefont {A.~M.}\ \bibnamefont
  {Black-Schaffer}},\ }\bibfield  {title} {\bibinfo {title} {Odd-frequency
  superconducting pairing in junctions with rashba spin-orbit coupling},\
  }\href {https://doi.org/10.1103/PhysRevB.98.075425} {\bibfield  {journal}
  {\bibinfo  {journal} {Phys. Rev. B}\ }\textbf {\bibinfo {volume} {98}},\
  \bibinfo {pages} {075425} (\bibinfo {year} {2018})}\BibitemShut {NoStop}%
\bibitem [{\citenamefont {Linder}\ \emph {et~al.}(2010)\citenamefont {Linder},
  \citenamefont {Sudb\o{}}, \citenamefont {Yokoyama}, \citenamefont {Grein},\
  and\ \citenamefont {Eschrig}}]{lind1}%
  \BibitemOpen
  \bibfield  {author} {\bibinfo {author} {\bibfnamefont {J.}~\bibnamefont
  {Linder}}, \bibinfo {author} {\bibfnamefont {A.}~\bibnamefont {Sudb\o{}}},
  \bibinfo {author} {\bibfnamefont {T.}~\bibnamefont {Yokoyama}}, \bibinfo
  {author} {\bibfnamefont {R.}~\bibnamefont {Grein}},\ and\ \bibinfo {author}
  {\bibfnamefont {M.}~\bibnamefont {Eschrig}},\ }\bibfield  {title} {\bibinfo
  {title} {Signature of odd-frequency pairing correlations induced by a
  magnetic interface},\ }\href {https://doi.org/10.1103/PhysRevB.81.214504}
  {\bibfield  {journal} {\bibinfo  {journal} {Phys. Rev. B}\ }\textbf {\bibinfo
  {volume} {81}},\ \bibinfo {pages} {214504} (\bibinfo {year}
  {2010})}\BibitemShut {NoStop}%
\bibitem [{\citenamefont {Linder}\ \emph {et~al.}(2009)\citenamefont {Linder},
  \citenamefont {Yokoyama}, \citenamefont {Sudb\o{}},\ and\ \citenamefont
  {Eschrig}}]{lindd2}%
  \BibitemOpen
  \bibfield  {author} {\bibinfo {author} {\bibfnamefont {J.}~\bibnamefont
  {Linder}}, \bibinfo {author} {\bibfnamefont {T.}~\bibnamefont {Yokoyama}},
  \bibinfo {author} {\bibfnamefont {A.}~\bibnamefont {Sudb\o{}}},\ and\
  \bibinfo {author} {\bibfnamefont {M.}~\bibnamefont {Eschrig}},\ }\bibfield
  {title} {\bibinfo {title} {Pairing symmetry conversion by spin-active
  interfaces in magnetic normal-metal--superconductor junctions},\ }\href
  {https://doi.org/10.1103/PhysRevLett.102.107008} {\bibfield  {journal}
  {\bibinfo  {journal} {Phys. Rev. Lett.}\ }\textbf {\bibinfo {volume} {102}},\
  \bibinfo {pages} {107008} (\bibinfo {year} {2009})}\BibitemShut {NoStop}%
\bibitem [{\citenamefont {Pal}\ and\ \citenamefont {Benjamin}(2021)}]{pal}%
  \BibitemOpen
  \bibfield  {author} {\bibinfo {author} {\bibfnamefont {S.}~\bibnamefont
  {Pal}}\ and\ \bibinfo {author} {\bibfnamefont {C.}~\bibnamefont {Benjamin}},\
  }\bibfield  {title} {\bibinfo {title} {Exciting odd-frequency equal-spin
  triplet correlations at metal-superconductor interfaces},\ }\href
  {https://doi.org/10.1103/PhysRevB.104.054519} {\bibfield  {journal} {\bibinfo
   {journal} {Phys. Rev. B}\ }\textbf {\bibinfo {volume} {104}},\ \bibinfo
  {pages} {054519} (\bibinfo {year} {2021})}\BibitemShut {NoStop}%
\bibitem [{\citenamefont {Tamura}\ \emph {et~al.}(2023)\citenamefont {Tamura},
  \citenamefont {Tanaka},\ and\ \citenamefont {Yokoyama}}]{tamu}%
  \BibitemOpen
  \bibfield  {author} {\bibinfo {author} {\bibfnamefont {S.}~\bibnamefont
  {Tamura}}, \bibinfo {author} {\bibfnamefont {Y.}~\bibnamefont {Tanaka}},\
  and\ \bibinfo {author} {\bibfnamefont {T.}~\bibnamefont {Yokoyama}},\
  }\bibfield  {title} {\bibinfo {title} {Generation of polarized spin-triplet
  cooper pairings by magnetic barriers in superconducting junctions},\ }\href
  {https://doi.org/10.1103/PhysRevB.107.054501} {\bibfield  {journal} {\bibinfo
   {journal} {Phys. Rev. B}\ }\textbf {\bibinfo {volume} {107}},\ \bibinfo
  {pages} {054501} (\bibinfo {year} {2023})}\BibitemShut {NoStop}%
\bibitem [{\citenamefont {Tanaka}\ and\ \citenamefont {Golubov}(2007)}]{aag}%
  \BibitemOpen
  \bibfield  {author} {\bibinfo {author} {\bibfnamefont {Y.}~\bibnamefont
  {Tanaka}}\ and\ \bibinfo {author} {\bibfnamefont {A.~A.}\ \bibnamefont
  {Golubov}},\ }\bibfield  {title} {\bibinfo {title} {Theory of the proximity
  effect in junctions with unconventional superconductors},\ }\href
  {https://doi.org/10.1103/PhysRevLett.98.037003} {\bibfield  {journal}
  {\bibinfo  {journal} {Phys. Rev. Lett.}\ }\textbf {\bibinfo {volume} {98}},\
  \bibinfo {pages} {037003} (\bibinfo {year} {2007})}\BibitemShut {NoStop}%
\bibitem [{\citenamefont {Eschrig}\ \emph {et~al.}(2007)\citenamefont
  {Eschrig}, \citenamefont {L{\"o}fwander}, \citenamefont {Champel},
  \citenamefont {Cuevas}, \citenamefont {Kopu},\ and\ \citenamefont
  {Sch{\"o}n}}]{lof}%
  \BibitemOpen
  \bibfield  {author} {\bibinfo {author} {\bibfnamefont {M.}~\bibnamefont
  {Eschrig}}, \bibinfo {author} {\bibfnamefont {T.}~\bibnamefont
  {L{\"o}fwander}}, \bibinfo {author} {\bibfnamefont {T.}~\bibnamefont
  {Champel}}, \bibinfo {author} {\bibfnamefont {J.}~\bibnamefont {Cuevas}},
  \bibinfo {author} {\bibfnamefont {J.}~\bibnamefont {Kopu}},\ and\ \bibinfo
  {author} {\bibfnamefont {G.}~\bibnamefont {Sch{\"o}n}},\ }\bibfield  {title}
  {\bibinfo {title} {Symmetries of pairing correlations in
  superconductor--ferromagnet nanostructures},\ }\href@noop {} {\bibfield
  {journal} {\bibinfo  {journal} {Journal of Low Temperature Physics}\ }\textbf
  {\bibinfo {volume} {147}},\ \bibinfo {pages} {457} (\bibinfo {year}
  {2007})}\BibitemShut {NoStop}%
\bibitem [{\citenamefont {Volkov}\ \emph {et~al.}(2003)\citenamefont {Volkov},
  \citenamefont {Bergeret},\ and\ \citenamefont {Efetov}}]{vol1}%
  \BibitemOpen
  \bibfield  {author} {\bibinfo {author} {\bibfnamefont {A.~F.}\ \bibnamefont
  {Volkov}}, \bibinfo {author} {\bibfnamefont {F.~S.}\ \bibnamefont
  {Bergeret}},\ and\ \bibinfo {author} {\bibfnamefont {K.~B.}\ \bibnamefont
  {Efetov}},\ }\bibfield  {title} {\bibinfo {title} {Odd triplet
  superconductivity in superconductor-ferromagnet multilayered structures},\
  }\href {https://doi.org/10.1103/PhysRevLett.90.117006} {\bibfield  {journal}
  {\bibinfo  {journal} {Phys. Rev. Lett.}\ }\textbf {\bibinfo {volume} {90}},\
  \bibinfo {pages} {117006} (\bibinfo {year} {2003})}\BibitemShut {NoStop}%
\bibitem [{\citenamefont {Fominov}\ \emph {et~al.}(2007)\citenamefont
  {Fominov}, \citenamefont {Volkov},\ and\ \citenamefont {Efetov}}]{fom}%
  \BibitemOpen
  \bibfield  {author} {\bibinfo {author} {\bibfnamefont {Y.~V.}\ \bibnamefont
  {Fominov}}, \bibinfo {author} {\bibfnamefont {A.~F.}\ \bibnamefont
  {Volkov}},\ and\ \bibinfo {author} {\bibfnamefont {K.~B.}\ \bibnamefont
  {Efetov}},\ }\bibfield  {title} {\bibinfo {title} {Josephson effect due to
  the long-range odd-frequency triplet superconductivity in
  $\mathrm{S}\mathrm{F}\mathrm{S}$ junctions with n\'eel domain walls},\ }\href
  {https://doi.org/10.1103/PhysRevB.75.104509} {\bibfield  {journal} {\bibinfo
  {journal} {Phys. Rev. B}\ }\textbf {\bibinfo {volume} {75}},\ \bibinfo
  {pages} {104509} (\bibinfo {year} {2007})}\BibitemShut {NoStop}%
\bibitem [{\citenamefont {Buzdin}(2005)}]{buz}%
  \BibitemOpen
  \bibfield  {author} {\bibinfo {author} {\bibfnamefont {A.~I.}\ \bibnamefont
  {Buzdin}},\ }\bibfield  {title} {\bibinfo {title} {Proximity effects in
  superconductor-ferromagnet heterostructures},\ }\href
  {https://doi.org/10.1103/RevModPhys.77.935} {\bibfield  {journal} {\bibinfo
  {journal} {Rev. Mod. Phys.}\ }\textbf {\bibinfo {volume} {77}},\ \bibinfo
  {pages} {935} (\bibinfo {year} {2005})}\BibitemShut {NoStop}%
\bibitem [{\citenamefont {Tsintzis}\ \emph {et~al.}(2019)\citenamefont
  {Tsintzis}, \citenamefont {Black-Schaffer},\ and\ \citenamefont
  {Cayao}}]{ats}%
  \BibitemOpen
  \bibfield  {author} {\bibinfo {author} {\bibfnamefont {A.}~\bibnamefont
  {Tsintzis}}, \bibinfo {author} {\bibfnamefont {A.~M.}\ \bibnamefont
  {Black-Schaffer}},\ and\ \bibinfo {author} {\bibfnamefont {J.}~\bibnamefont
  {Cayao}},\ }\bibfield  {title} {\bibinfo {title} {Odd-frequency
  superconducting pairing in kitaev-based junctions},\ }\href
  {https://doi.org/10.1103/PhysRevB.100.115433} {\bibfield  {journal} {\bibinfo
   {journal} {Phys. Rev. B}\ }\textbf {\bibinfo {volume} {100}},\ \bibinfo
  {pages} {115433} (\bibinfo {year} {2019})}\BibitemShut {NoStop}%
\bibitem [{\citenamefont {Asano}\ and\ \citenamefont {Tanaka}(2013)}]{asan}%
  \BibitemOpen
  \bibfield  {author} {\bibinfo {author} {\bibfnamefont {Y.}~\bibnamefont
  {Asano}}\ and\ \bibinfo {author} {\bibfnamefont {Y.}~\bibnamefont {Tanaka}},\
  }\bibfield  {title} {\bibinfo {title} {Majorana fermions and odd-frequency
  cooper pairs in a normal-metal nanowire proximity-coupled to a topological
  superconductor},\ }\href {https://doi.org/10.1103/PhysRevB.87.104513}
  {\bibfield  {journal} {\bibinfo  {journal} {Phys. Rev. B}\ }\textbf {\bibinfo
  {volume} {87}},\ \bibinfo {pages} {104513} (\bibinfo {year}
  {2013})}\BibitemShut {NoStop}%
\bibitem [{\citenamefont {Kuzmanovski}\ and\ \citenamefont
  {Black-Schaffer}(2017)}]{dku}%
  \BibitemOpen
  \bibfield  {author} {\bibinfo {author} {\bibfnamefont {D.}~\bibnamefont
  {Kuzmanovski}}\ and\ \bibinfo {author} {\bibfnamefont {A.~M.}\ \bibnamefont
  {Black-Schaffer}},\ }\bibfield  {title} {\bibinfo {title} {Multiple
  odd-frequency superconducting states in buckled quantum spin hall insulators
  with time-reversal symmetry},\ }\href
  {https://doi.org/10.1103/PhysRevB.96.174509} {\bibfield  {journal} {\bibinfo
  {journal} {Phys. Rev. B}\ }\textbf {\bibinfo {volume} {96}},\ \bibinfo
  {pages} {174509} (\bibinfo {year} {2017})}\BibitemShut {NoStop}%
\bibitem [{\citenamefont {Fleckenstein}\ \emph {et~al.}(2018)\citenamefont
  {Fleckenstein}, \citenamefont {Ziani},\ and\ \citenamefont
  {Trauzettel}}]{CFL}%
  \BibitemOpen
  \bibfield  {author} {\bibinfo {author} {\bibfnamefont {C.}~\bibnamefont
  {Fleckenstein}}, \bibinfo {author} {\bibfnamefont {N.~T.}\ \bibnamefont
  {Ziani}},\ and\ \bibinfo {author} {\bibfnamefont {B.}~\bibnamefont
  {Trauzettel}},\ }\bibfield  {title} {\bibinfo {title} {Conductance signatures
  of odd-frequency superconductivity in quantum spin hall systems using a
  quantum point contact},\ }\href {https://doi.org/10.1103/PhysRevB.97.134523}
  {\bibfield  {journal} {\bibinfo  {journal} {Phys. Rev. B}\ }\textbf {\bibinfo
  {volume} {97}},\ \bibinfo {pages} {134523} (\bibinfo {year}
  {2018})}\BibitemShut {NoStop}%
\bibitem [{\citenamefont {Tanaka}\ and\ \citenamefont
  {Tamura}(2021)}]{tanaka2021theory}%
  \BibitemOpen
  \bibfield  {author} {\bibinfo {author} {\bibfnamefont {Y.}~\bibnamefont
  {Tanaka}}\ and\ \bibinfo {author} {\bibfnamefont {S.}~\bibnamefont
  {Tamura}},\ }\bibfield  {title} {\bibinfo {title} {Theory of surface andreev
  bound states and odd-frequency pairing in superconductor junctions},\
  }\href@noop {} {\bibfield  {journal} {\bibinfo  {journal} {Journal of
  Superconductivity and Novel Magnetism}\ }\textbf {\bibinfo {volume} {34}},\
  \bibinfo {pages} {1677} (\bibinfo {year} {2021})}\BibitemShut {NoStop}%
\bibitem [{\citenamefont {Tanaka}\ \emph
  {et~al.}(2007{\natexlab{b}})\citenamefont {Tanaka}, \citenamefont {Golubov},
  \citenamefont {Kashiwaya},\ and\ \citenamefont
  {Ueda}}]{PhysRevLett.99.037005}%
  \BibitemOpen
  \bibfield  {author} {\bibinfo {author} {\bibfnamefont {Y.}~\bibnamefont
  {Tanaka}}, \bibinfo {author} {\bibfnamefont {A.~A.}\ \bibnamefont {Golubov}},
  \bibinfo {author} {\bibfnamefont {S.}~\bibnamefont {Kashiwaya}},\ and\
  \bibinfo {author} {\bibfnamefont {M.}~\bibnamefont {Ueda}},\ }\bibfield
  {title} {\bibinfo {title} {Anomalous josephson effect between even- and
  odd-frequency superconductors},\ }\href
  {https://doi.org/10.1103/PhysRevLett.99.037005} {\bibfield  {journal}
  {\bibinfo  {journal} {Phys. Rev. Lett.}\ }\textbf {\bibinfo {volume} {99}},\
  \bibinfo {pages} {037005} (\bibinfo {year} {2007}{\natexlab{b}})}\BibitemShut
  {NoStop}%
\bibitem [{\citenamefont {Triola}\ and\ \citenamefont
  {Balatsky}(2016)}]{trio1}%
  \BibitemOpen
  \bibfield  {author} {\bibinfo {author} {\bibfnamefont {C.}~\bibnamefont
  {Triola}}\ and\ \bibinfo {author} {\bibfnamefont {A.~V.}\ \bibnamefont
  {Balatsky}},\ }\bibfield  {title} {\bibinfo {title} {Odd-frequency
  superconductivity in driven systems},\ }\href
  {https://doi.org/10.1103/PhysRevB.94.094518} {\bibfield  {journal} {\bibinfo
  {journal} {Phys. Rev. B}\ }\textbf {\bibinfo {volume} {94}},\ \bibinfo
  {pages} {094518} (\bibinfo {year} {2016})}\BibitemShut {NoStop}%
\bibitem [{\citenamefont {Triola}\ and\ \citenamefont
  {Balatsky}(2017)}]{trio2}%
  \BibitemOpen
  \bibfield  {author} {\bibinfo {author} {\bibfnamefont {C.}~\bibnamefont
  {Triola}}\ and\ \bibinfo {author} {\bibfnamefont {A.~V.}\ \bibnamefont
  {Balatsky}},\ }\bibfield  {title} {\bibinfo {title} {Pair symmetry conversion
  in driven multiband superconductors},\ }\href
  {https://doi.org/10.1103/PhysRevB.95.224518} {\bibfield  {journal} {\bibinfo
  {journal} {Phys. Rev. B}\ }\textbf {\bibinfo {volume} {95}},\ \bibinfo
  {pages} {224518} (\bibinfo {year} {2017})}\BibitemShut {NoStop}%
\bibitem [{\citenamefont {Cayao}\ \emph {et~al.}(2022)\citenamefont {Cayao},
  \citenamefont {Dutta}, \citenamefont {Burset},\ and\ \citenamefont
  {Black-Schaffer}}]{jca}%
  \BibitemOpen
  \bibfield  {author} {\bibinfo {author} {\bibfnamefont {J.}~\bibnamefont
  {Cayao}}, \bibinfo {author} {\bibfnamefont {P.}~\bibnamefont {Dutta}},
  \bibinfo {author} {\bibfnamefont {P.}~\bibnamefont {Burset}},\ and\ \bibinfo
  {author} {\bibfnamefont {A.~M.}\ \bibnamefont {Black-Schaffer}},\ }\bibfield
  {title} {\bibinfo {title} {Phase-tunable electron transport assisted by
  odd-frequency cooper pairs in topological josephson junctions},\ }\href
  {https://doi.org/10.1103/PhysRevB.106.L100502} {\bibfield  {journal}
  {\bibinfo  {journal} {Phys. Rev. B}\ }\textbf {\bibinfo {volume} {106}},\
  \bibinfo {pages} {L100502} (\bibinfo {year} {2022})}\BibitemShut {NoStop}%
\bibitem [{\citenamefont {Dutta}\ and\ \citenamefont
  {Black-Schaffer}(2019)}]{pdut}%
  \BibitemOpen
  \bibfield  {author} {\bibinfo {author} {\bibfnamefont {P.}~\bibnamefont
  {Dutta}}\ and\ \bibinfo {author} {\bibfnamefont {A.~M.}\ \bibnamefont
  {Black-Schaffer}},\ }\bibfield  {title} {\bibinfo {title} {Signature of
  odd-frequency equal-spin triplet pairing in the josephson current on the
  surface of weyl nodal loop semimetals},\ }\href
  {https://doi.org/10.1103/PhysRevB.100.104511} {\bibfield  {journal} {\bibinfo
   {journal} {Phys. Rev. B}\ }\textbf {\bibinfo {volume} {100}},\ \bibinfo
  {pages} {104511} (\bibinfo {year} {2019})}\BibitemShut {NoStop}%
\bibitem [{\citenamefont {Seoane~Souto}\ \emph {et~al.}(2020)\citenamefont
  {Seoane~Souto}, \citenamefont {Kuzmanovski},\ and\ \citenamefont
  {Balatsky}}]{rse}%
  \BibitemOpen
  \bibfield  {author} {\bibinfo {author} {\bibfnamefont {R.}~\bibnamefont
  {Seoane~Souto}}, \bibinfo {author} {\bibfnamefont {D.}~\bibnamefont
  {Kuzmanovski}},\ and\ \bibinfo {author} {\bibfnamefont {A.~V.}\ \bibnamefont
  {Balatsky}},\ }\bibfield  {title} {\bibinfo {title} {Signatures of
  odd-frequency pairing in the josephson junction current noise},\ }\href
  {https://doi.org/10.1103/PhysRevResearch.2.043193} {\bibfield  {journal}
  {\bibinfo  {journal} {Phys. Rev. Res.}\ }\textbf {\bibinfo {volume} {2}},\
  \bibinfo {pages} {043193} (\bibinfo {year} {2020})}\BibitemShut {NoStop}%
\bibitem [{\citenamefont {Khaire}\ \emph {et~al.}(2010)\citenamefont {Khaire},
  \citenamefont {Khasawneh}, \citenamefont {Pratt},\ and\ \citenamefont
  {Birge}}]{kha}%
  \BibitemOpen
  \bibfield  {author} {\bibinfo {author} {\bibfnamefont {T.~S.}\ \bibnamefont
  {Khaire}}, \bibinfo {author} {\bibfnamefont {M.~A.}\ \bibnamefont
  {Khasawneh}}, \bibinfo {author} {\bibfnamefont {W.~P.}\ \bibnamefont
  {Pratt}},\ and\ \bibinfo {author} {\bibfnamefont {N.~O.}\ \bibnamefont
  {Birge}},\ }\bibfield  {title} {\bibinfo {title} {Observation of spin-triplet
  superconductivity in co-based josephson junctions},\ }\href
  {https://doi.org/10.1103/PhysRevLett.104.137002} {\bibfield  {journal}
  {\bibinfo  {journal} {Phys. Rev. Lett.}\ }\textbf {\bibinfo {volume} {104}},\
  \bibinfo {pages} {137002} (\bibinfo {year} {2010})}\BibitemShut {NoStop}%
\bibitem [{\citenamefont {Di~Bernardo}\ \emph
  {et~al.}(2015{\natexlab{b}})\citenamefont {Di~Bernardo}, \citenamefont
  {Diesch}, \citenamefont {Gu}, \citenamefont {Linder}, \citenamefont
  {Divitini}, \citenamefont {Ducati}, \citenamefont {Scheer}, \citenamefont
  {Blamire},\ and\ \citenamefont {Robinson}}]{adb}%
  \BibitemOpen
  \bibfield  {author} {\bibinfo {author} {\bibfnamefont {A.}~\bibnamefont
  {Di~Bernardo}}, \bibinfo {author} {\bibfnamefont {S.}~\bibnamefont {Diesch}},
  \bibinfo {author} {\bibfnamefont {Y.}~\bibnamefont {Gu}}, \bibinfo {author}
  {\bibfnamefont {J.}~\bibnamefont {Linder}}, \bibinfo {author} {\bibfnamefont
  {G.}~\bibnamefont {Divitini}}, \bibinfo {author} {\bibfnamefont
  {C.}~\bibnamefont {Ducati}}, \bibinfo {author} {\bibfnamefont
  {E.}~\bibnamefont {Scheer}}, \bibinfo {author} {\bibfnamefont {M.~G.}\
  \bibnamefont {Blamire}},\ and\ \bibinfo {author} {\bibfnamefont {J.~W.}\
  \bibnamefont {Robinson}},\ }\bibfield  {title} {\bibinfo {title} {Signature
  of magnetic-dependent gapless odd frequency states at
  superconductor/ferromagnet interfaces},\ }\href@noop {} {\bibfield  {journal}
  {\bibinfo  {journal} {Nature communications}\ }\textbf {\bibinfo {volume}
  {6}},\ \bibinfo {pages} {8053} (\bibinfo {year}
  {2015}{\natexlab{b}})}\BibitemShut {NoStop}%
\bibitem [{\citenamefont {Hart}\ \emph {et~al.}(2014)\citenamefont {Hart},
  \citenamefont {Ren}, \citenamefont {Wagner}, \citenamefont {Leubner},
  \citenamefont {M{\"u}hlbauer}, \citenamefont {Br{\"u}ne}, \citenamefont
  {Buhmann}, \citenamefont {Molenkamp},\ and\ \citenamefont
  {Yacoby}}]{yacoby_NP_10_638}%
  \BibitemOpen
  \bibfield  {author} {\bibinfo {author} {\bibfnamefont {S.}~\bibnamefont
  {Hart}}, \bibinfo {author} {\bibfnamefont {H.}~\bibnamefont {Ren}}, \bibinfo
  {author} {\bibfnamefont {T.}~\bibnamefont {Wagner}}, \bibinfo {author}
  {\bibfnamefont {P.}~\bibnamefont {Leubner}}, \bibinfo {author} {\bibfnamefont
  {M.}~\bibnamefont {M{\"u}hlbauer}}, \bibinfo {author} {\bibfnamefont
  {C.}~\bibnamefont {Br{\"u}ne}}, \bibinfo {author} {\bibfnamefont
  {H.}~\bibnamefont {Buhmann}}, \bibinfo {author} {\bibfnamefont {L.~W.}\
  \bibnamefont {Molenkamp}},\ and\ \bibinfo {author} {\bibfnamefont
  {A.}~\bibnamefont {Yacoby}},\ }\bibfield  {title} {\bibinfo {title} {Induced
  superconductivity in the quantum spin hall edge},\ }\href@noop {} {\bibfield
  {journal} {\bibinfo  {journal} {Nature Physics}\ }\textbf {\bibinfo {volume}
  {10}},\ \bibinfo {pages} {638} (\bibinfo {year} {2014})}\BibitemShut
  {NoStop}%
\bibitem [{\citenamefont {Br{\"u}ne}\ \emph {et~al.}(2012)\citenamefont
  {Br{\"u}ne}, \citenamefont {Roth}, \citenamefont {Buhmann}, \citenamefont
  {Hankiewicz}, \citenamefont {Molenkamp}, \citenamefont {Maciejko},
  \citenamefont {Qi},\ and\ \citenamefont {Zhang}}]{molenkamp_NP_8_485}%
  \BibitemOpen
  \bibfield  {author} {\bibinfo {author} {\bibfnamefont {C.}~\bibnamefont
  {Br{\"u}ne}}, \bibinfo {author} {\bibfnamefont {A.}~\bibnamefont {Roth}},
  \bibinfo {author} {\bibfnamefont {H.}~\bibnamefont {Buhmann}}, \bibinfo
  {author} {\bibfnamefont {E.~M.}\ \bibnamefont {Hankiewicz}}, \bibinfo
  {author} {\bibfnamefont {L.~W.}\ \bibnamefont {Molenkamp}}, \bibinfo {author}
  {\bibfnamefont {J.}~\bibnamefont {Maciejko}}, \bibinfo {author}
  {\bibfnamefont {X.-L.}\ \bibnamefont {Qi}},\ and\ \bibinfo {author}
  {\bibfnamefont {S.-C.}\ \bibnamefont {Zhang}},\ }\bibfield  {title} {\bibinfo
  {title} {Spin polarization of the quantum spin hall edge states},\
  }\href@noop {} {\bibfield  {journal} {\bibinfo  {journal} {Nature Physics}\
  }\textbf {\bibinfo {volume} {8}},\ \bibinfo {pages} {485} (\bibinfo {year}
  {2012})}\BibitemShut {NoStop}%
\bibitem [{\citenamefont {Das}\ and\ \citenamefont {Rao}(2011)}]{suma}%
  \BibitemOpen
  \bibfield  {author} {\bibinfo {author} {\bibfnamefont {S.}~\bibnamefont
  {Das}}\ and\ \bibinfo {author} {\bibfnamefont {S.}~\bibnamefont {Rao}},\
  }\bibfield  {title} {\bibinfo {title} {Spin-polarized scanning-tunneling
  probe for helical luttinger liquids},\ }\href
  {https://doi.org/10.1103/PhysRevLett.106.236403} {\bibfield  {journal}
  {\bibinfo  {journal} {Phys. Rev. Lett.}\ }\textbf {\bibinfo {volume} {106}},\
  \bibinfo {pages} {236403} (\bibinfo {year} {2011})}\BibitemShut {NoStop}%
\bibitem [{\citenamefont {Fu}\ and\ \citenamefont {Kane}(2009)}]{lu}%
  \BibitemOpen
  \bibfield  {author} {\bibinfo {author} {\bibfnamefont {L.}~\bibnamefont
  {Fu}}\ and\ \bibinfo {author} {\bibfnamefont {C.~L.}\ \bibnamefont {Kane}},\
  }\bibfield  {title} {\bibinfo {title} {Josephson current and noise at a
  superconductor/quantum-spin-hall-insulator/superconductor junction},\ }\href
  {https://doi.org/10.1103/PhysRevB.79.161408} {\bibfield  {journal} {\bibinfo
  {journal} {Phys. Rev. B}\ }\textbf {\bibinfo {volume} {79}},\ \bibinfo
  {pages} {161408} (\bibinfo {year} {2009})}\BibitemShut {NoStop}%
\bibitem [{\citenamefont {Fu}\ and\ \citenamefont {Kane}(2008)}]{lu1}%
  \BibitemOpen
  \bibfield  {author} {\bibinfo {author} {\bibfnamefont {L.}~\bibnamefont
  {Fu}}\ and\ \bibinfo {author} {\bibfnamefont {C.~L.}\ \bibnamefont {Kane}},\
  }\bibfield  {title} {\bibinfo {title} {Superconducting proximity effect and
  majorana fermions at the surface of a topological insulator},\ }\href
  {https://doi.org/10.1103/PhysRevLett.100.096407} {\bibfield  {journal}
  {\bibinfo  {journal} {Phys. Rev. Lett.}\ }\textbf {\bibinfo {volume} {100}},\
  \bibinfo {pages} {096407} (\bibinfo {year} {2008})}\BibitemShut {NoStop}%
\bibitem [{\citenamefont {Kane}\ and\ \citenamefont {Mele}(2005)}]{clk}%
  \BibitemOpen
  \bibfield  {author} {\bibinfo {author} {\bibfnamefont {C.~L.}\ \bibnamefont
  {Kane}}\ and\ \bibinfo {author} {\bibfnamefont {E.~J.}\ \bibnamefont
  {Mele}},\ }\bibfield  {title} {\bibinfo {title} {Quantum spin hall effect in
  graphene},\ }\href {https://doi.org/10.1103/PhysRevLett.95.226801} {\bibfield
   {journal} {\bibinfo  {journal} {Phys. Rev. Lett.}\ }\textbf {\bibinfo
  {volume} {95}},\ \bibinfo {pages} {226801} (\bibinfo {year}
  {2005})}\BibitemShut {NoStop}%
\bibitem [{\citenamefont {Calzona}\ and\ \citenamefont
  {Trauzettel}(2019)}]{call}%
  \BibitemOpen
  \bibfield  {author} {\bibinfo {author} {\bibfnamefont {A.}~\bibnamefont
  {Calzona}}\ and\ \bibinfo {author} {\bibfnamefont {B.}~\bibnamefont
  {Trauzettel}},\ }\bibfield  {title} {\bibinfo {title} {Moving majorana bound
  states between distinct helical edges across a quantum point contact},\
  }\href {https://doi.org/10.1103/PhysRevResearch.1.033212} {\bibfield
  {journal} {\bibinfo  {journal} {Phys. Rev. Res.}\ }\textbf {\bibinfo {volume}
  {1}},\ \bibinfo {pages} {033212} (\bibinfo {year} {2019})}\BibitemShut
  {NoStop}%
\bibitem [{\citenamefont {Keidel}\ \emph {et~al.}(2020)\citenamefont {Keidel},
  \citenamefont {Hwang}, \citenamefont {Trauzettel}, \citenamefont {Sothmann},\
  and\ \citenamefont {Burset}}]{keidel_PRR_2_022019}%
  \BibitemOpen
  \bibfield  {author} {\bibinfo {author} {\bibfnamefont {F.}~\bibnamefont
  {Keidel}}, \bibinfo {author} {\bibfnamefont {S.-Y.}\ \bibnamefont {Hwang}},
  \bibinfo {author} {\bibfnamefont {B.}~\bibnamefont {Trauzettel}}, \bibinfo
  {author} {\bibfnamefont {B.}~\bibnamefont {Sothmann}},\ and\ \bibinfo
  {author} {\bibfnamefont {P.}~\bibnamefont {Burset}},\ }\bibfield  {title}
  {\bibinfo {title} {On-demand thermoelectric generation of equal-spin cooper
  pairs},\ }\href {https://doi.org/10.1103/PhysRevResearch.2.022019} {\bibfield
   {journal} {\bibinfo  {journal} {Phys. Rev. Res.}\ }\textbf {\bibinfo
  {volume} {2}},\ \bibinfo {pages} {022019} (\bibinfo {year}
  {2020})}\BibitemShut {NoStop}%
\bibitem [{\citenamefont {Mukhopadhyay}\ and\ \citenamefont
  {Das}(2021)}]{aab1}%
  \BibitemOpen
  \bibfield  {author} {\bibinfo {author} {\bibfnamefont {A.}~\bibnamefont
  {Mukhopadhyay}}\ and\ \bibinfo {author} {\bibfnamefont {S.}~\bibnamefont
  {Das}},\ }\bibfield  {title} {\bibinfo {title} {Thermal signature of the
  majorana fermion in a josephson junction},\ }\href
  {https://doi.org/10.1103/PhysRevB.103.144502} {\bibfield  {journal} {\bibinfo
   {journal} {Phys. Rev. B}\ }\textbf {\bibinfo {volume} {103}},\ \bibinfo
  {pages} {144502} (\bibinfo {year} {2021})}\BibitemShut {NoStop}%
\bibitem [{\citenamefont {Mukhopadhyay}\ and\ \citenamefont
  {Das}(2022)}]{aab2}%
  \BibitemOpen
  \bibfield  {author} {\bibinfo {author} {\bibfnamefont {A.}~\bibnamefont
  {Mukhopadhyay}}\ and\ \bibinfo {author} {\bibfnamefont {S.}~\bibnamefont
  {Das}},\ }\bibfield  {title} {\bibinfo {title} {Thermal bias induced charge
  current in a josephson junction: From ballistic to disordered},\ }\href
  {https://doi.org/10.1103/PhysRevB.106.075421} {\bibfield  {journal} {\bibinfo
   {journal} {Phys. Rev. B}\ }\textbf {\bibinfo {volume} {106}},\ \bibinfo
  {pages} {075421} (\bibinfo {year} {2022})}\BibitemShut {NoStop}%
\bibitem [{\citenamefont {Wadhawan}\ \emph {et~al.}(2018)\citenamefont
  {Wadhawan}, \citenamefont {Roychowdhury}, \citenamefont {Mehta},\ and\
  \citenamefont {Das}}]{wad}%
  \BibitemOpen
  \bibfield  {author} {\bibinfo {author} {\bibfnamefont {D.}~\bibnamefont
  {Wadhawan}}, \bibinfo {author} {\bibfnamefont {K.}~\bibnamefont
  {Roychowdhury}}, \bibinfo {author} {\bibfnamefont {P.}~\bibnamefont
  {Mehta}},\ and\ \bibinfo {author} {\bibfnamefont {S.}~\bibnamefont {Das}},\
  }\bibfield  {title} {\bibinfo {title} {Multielectron geometric phase in
  intensity interferometry},\ }\href
  {https://doi.org/10.1103/PhysRevB.98.155113} {\bibfield  {journal} {\bibinfo
  {journal} {Phys. Rev. B}\ }\textbf {\bibinfo {volume} {98}},\ \bibinfo
  {pages} {155113} (\bibinfo {year} {2018})}\BibitemShut {NoStop}%
\bibitem [{Note1()}]{Note1}%
  \BibitemOpen
  \bibinfo {note} {The azimuthal angle is of no consequence, e.g., it can also
  be assumed in the $z-y$ plane}\BibitemShut {NoStop}%
\bibitem [{\citenamefont {Pershoguba}\ \emph {et~al.}(2019)\citenamefont
  {Pershoguba}, \citenamefont {Veness},\ and\ \citenamefont {Glazman}}]{glaz}%
  \BibitemOpen
  \bibfield  {author} {\bibinfo {author} {\bibfnamefont {S.~S.}\ \bibnamefont
  {Pershoguba}}, \bibinfo {author} {\bibfnamefont {T.}~\bibnamefont {Veness}},\
  and\ \bibinfo {author} {\bibfnamefont {L.~I.}\ \bibnamefont {Glazman}},\
  }\bibfield  {title} {\bibinfo {title} {Landauer formula for a superconducting
  quantum point contact},\ }\href
  {https://doi.org/10.1103/PhysRevLett.123.067001} {\bibfield  {journal}
  {\bibinfo  {journal} {Phys. Rev. Lett.}\ }\textbf {\bibinfo {volume} {123}},\
  \bibinfo {pages} {067001} (\bibinfo {year} {2019})}\BibitemShut {NoStop}%
\bibitem [{\citenamefont {Groth}\ \emph {et~al.}(2014)\citenamefont {Groth},
  \citenamefont {Wimmer}, \citenamefont {Akhmerov},\ and\ \citenamefont
  {Waintal}}]{groth2014kwant}%
  \BibitemOpen
  \bibfield  {author} {\bibinfo {author} {\bibfnamefont {C.~W.}\ \bibnamefont
  {Groth}}, \bibinfo {author} {\bibfnamefont {M.}~\bibnamefont {Wimmer}},
  \bibinfo {author} {\bibfnamefont {A.~R.}\ \bibnamefont {Akhmerov}},\ and\
  \bibinfo {author} {\bibfnamefont {X.}~\bibnamefont {Waintal}},\ }\bibfield
  {title} {\bibinfo {title} {Kwant: a software package for quantum transport},\
  }\href@noop {} {\bibfield  {journal} {\bibinfo  {journal} {New Journal of
  Physics}\ }\textbf {\bibinfo {volume} {16}},\ \bibinfo {pages} {063065}
  (\bibinfo {year} {2014})}\BibitemShut {NoStop}%
\bibitem [{\citenamefont {Bernevig}\ \emph {et~al.}(2006)\citenamefont
  {Bernevig}, \citenamefont {Hughes},\ and\ \citenamefont
  {Zhang}}]{BHZ_SC_314_1757}%
  \BibitemOpen
  \bibfield  {author} {\bibinfo {author} {\bibfnamefont {B.~A.}\ \bibnamefont
  {Bernevig}}, \bibinfo {author} {\bibfnamefont {T.~L.}\ \bibnamefont
  {Hughes}},\ and\ \bibinfo {author} {\bibfnamefont {S.-C.}\ \bibnamefont
  {Zhang}},\ }\bibfield  {title} {\bibinfo {title} {Quantum spin hall effect
  and topological phase transition in hgte quantum wells},\ }\href
  {https://doi.org/10.1126/science.1133734} {\bibfield  {journal} {\bibinfo
  {journal} {Science}\ }\textbf {\bibinfo {volume} {314}},\ \bibinfo {pages}
  {1757} (\bibinfo {year} {2006})},\ \Eprint
  {https://arxiv.org/abs/https://www.science.org/doi/pdf/10.1126/science.1133734}
  {https://www.science.org/doi/pdf/10.1126/science.1133734} \BibitemShut
  {NoStop}%
\bibitem [{\citenamefont {Alicea}(2012)}]{alicea_RPP_75_076501}%
  \BibitemOpen
  \bibfield  {author} {\bibinfo {author} {\bibfnamefont {J.}~\bibnamefont
  {Alicea}},\ }\bibfield  {title} {\bibinfo {title} {New directions in the
  pursuit of majorana fermions in solid state systems},\ }\href@noop {}
  {\bibfield  {journal} {\bibinfo  {journal} {Reports on progress in physics}\
  }\textbf {\bibinfo {volume} {75}},\ \bibinfo {pages} {076501} (\bibinfo
  {year} {2012})}\BibitemShut {NoStop}%
\bibitem [{\citenamefont {Li}\ \emph {et~al.}(2013)\citenamefont {Li},
  \citenamefont {Sheng}, \citenamefont {Shen}, \citenamefont {Shao},
  \citenamefont {Wang}, \citenamefont {Sheng},\ and\ \citenamefont
  {Xing}}]{li_PRL_110_266802}%
  \BibitemOpen
  \bibfield  {author} {\bibinfo {author} {\bibfnamefont {H.}~\bibnamefont
  {Li}}, \bibinfo {author} {\bibfnamefont {L.}~\bibnamefont {Sheng}}, \bibinfo
  {author} {\bibfnamefont {R.}~\bibnamefont {Shen}}, \bibinfo {author}
  {\bibfnamefont {L.~B.}\ \bibnamefont {Shao}}, \bibinfo {author}
  {\bibfnamefont {B.}~\bibnamefont {Wang}}, \bibinfo {author} {\bibfnamefont
  {D.~N.}\ \bibnamefont {Sheng}},\ and\ \bibinfo {author} {\bibfnamefont
  {D.~Y.}\ \bibnamefont {Xing}},\ }\bibfield  {title} {\bibinfo {title}
  {Stabilization of the quantum spin hall effect by designed removal of
  time-reversal symmetry of edge states},\ }\href
  {https://doi.org/10.1103/PhysRevLett.110.266802} {\bibfield  {journal}
  {\bibinfo  {journal} {Phys. Rev. Lett.}\ }\textbf {\bibinfo {volume} {110}},\
  \bibinfo {pages} {266802} (\bibinfo {year} {2013})}\BibitemShut {NoStop}%
\bibitem [{\citenamefont {Adak}\ \emph
  {et~al.}(2022{\natexlab{a}})\citenamefont {Adak}, \citenamefont
  {Roychowdhury},\ and\ \citenamefont {Das}}]{adak_PhysicaE_139_115125}%
  \BibitemOpen
  \bibfield  {author} {\bibinfo {author} {\bibfnamefont {V.}~\bibnamefont
  {Adak}}, \bibinfo {author} {\bibfnamefont {K.}~\bibnamefont {Roychowdhury}},\
  and\ \bibinfo {author} {\bibfnamefont {S.}~\bibnamefont {Das}},\ }\bibfield
  {title} {\bibinfo {title} {Spin-polarized voltage probes for helical edge
  state: A model study},\ }\href
  {https://doi.org/https://doi.org/10.1016/j.physe.2021.115125} {\bibfield
  {journal} {\bibinfo  {journal} {Physica E: Low-dimensional Systems and
  Nanostructures}\ }\textbf {\bibinfo {volume} {139}},\ \bibinfo {pages}
  {115125} (\bibinfo {year} {2022}{\natexlab{a}})}\BibitemShut {NoStop}%
\bibitem [{\citenamefont {Adak}\ \emph
  {et~al.}(2022{\natexlab{b}})\citenamefont {Adak}, \citenamefont
  {Mukhopadhyay}, \citenamefont {De}, \citenamefont {Khanna}, \citenamefont
  {Rao},\ and\ \citenamefont {Das}}]{adak_PRB_106_045422}%
  \BibitemOpen
  \bibfield  {author} {\bibinfo {author} {\bibfnamefont {V.}~\bibnamefont
  {Adak}}, \bibinfo {author} {\bibfnamefont {A.}~\bibnamefont {Mukhopadhyay}},
  \bibinfo {author} {\bibfnamefont {S.~J.}\ \bibnamefont {De}}, \bibinfo
  {author} {\bibfnamefont {U.}~\bibnamefont {Khanna}}, \bibinfo {author}
  {\bibfnamefont {S.}~\bibnamefont {Rao}},\ and\ \bibinfo {author}
  {\bibfnamefont {S.}~\bibnamefont {Das}},\ }\bibfield  {title} {\bibinfo
  {title} {Chiral detection of majorana bound states at the edge of a quantum
  spin hall insulator},\ }\href {https://doi.org/10.1103/PhysRevB.106.045422}
  {\bibfield  {journal} {\bibinfo  {journal} {Phys. Rev. B}\ }\textbf {\bibinfo
  {volume} {106}},\ \bibinfo {pages} {045422} (\bibinfo {year}
  {2022}{\natexlab{b}})}\BibitemShut {NoStop}%
\end{thebibliography}%


\begin{thebibliography}{4}%
\makeatletter
\providecommand \@ifxundefined [1]{%
 \@ifx{#1\undefined}
}%
\providecommand \@ifnum [1]{%
 \ifnum #1\expandafter \@firstoftwo
 \else \expandafter \@secondoftwo
 \fi
}%
\providecommand \@ifx [1]{%
 \ifx #1\expandafter \@firstoftwo
 \else \expandafter \@secondoftwo
 \fi
}%
\providecommand \natexlab [1]{#1}%
\providecommand \enquote  [1]{``#1''}%
\providecommand \bibnamefont  [1]{#1}%
\providecommand \bibfnamefont [1]{#1}%
\providecommand \citenamefont [1]{#1}%
\providecommand \href@noop [0]{\@secondoftwo}%
\providecommand \href [0]{\begingroup \@sanitize@url \@href}%
\providecommand \@href[1]{\@@startlink{#1}\@@href}%
\providecommand \@@href[1]{\endgroup#1\@@endlink}%
\providecommand \@sanitize@url [0]{\catcode `\\12\catcode `\$12\catcode
  `\&12\catcode `\#12\catcode `\^12\catcode `\_12\catcode `\%12\relax}%
\providecommand \@@startlink[1]{}%
\providecommand \@@endlink[0]{}%
\providecommand \url  [0]{\begingroup\@sanitize@url \@url }%
\providecommand \@url [1]{\endgroup\@href {#1}{\urlprefix }}%
\providecommand \urlprefix  [0]{URL }%
\providecommand \Eprint [0]{\href }%
\providecommand \doibase [0]{https://doi.org/}%
\providecommand \selectlanguage [0]{\@gobble}%
\providecommand \bibinfo  [0]{\@secondoftwo}%
\providecommand \bibfield  [0]{\@secondoftwo}%
\providecommand \translation [1]{[#1]}%
\providecommand \BibitemOpen [0]{}%
\providecommand \bibitemStop [0]{}%
\providecommand \bibitemNoStop [0]{.\EOS\space}%
\providecommand \EOS [0]{\spacefactor3000\relax}%
\providecommand \BibitemShut  [1]{\csname bibitem#1\endcsname}%
\let\auto@bib@innerbib\@empty
\bibitem [{\citenamefont {Wadhawan}\ \emph {et~al.}(2018)\citenamefont
  {Wadhawan}, \citenamefont {Roychowdhury}, \citenamefont {Mehta},\ and\
  \citenamefont {Das}}]{wad}%
  \BibitemOpen
  \bibfield  {author} {\bibinfo {author} {\bibfnamefont {D.}~\bibnamefont
  {Wadhawan}}, \bibinfo {author} {\bibfnamefont {K.}~\bibnamefont
  {Roychowdhury}}, \bibinfo {author} {\bibfnamefont {P.}~\bibnamefont
  {Mehta}},\ and\ \bibinfo {author} {\bibfnamefont {S.}~\bibnamefont {Das}},\
  }\bibfield  {title} {\bibinfo {title} {Multielectron geometric phase in
  intensity interferometry},\ }\href
  {https://doi.org/10.1103/PhysRevB.98.155113} {\bibfield  {journal} {\bibinfo
  {journal} {Phys. Rev. B}\ }\textbf {\bibinfo {volume} {98}},\ \bibinfo
  {pages} {155113} (\bibinfo {year} {2018})}\BibitemShut {NoStop}%
\bibitem [{\citenamefont {Keidel}\ \emph {et~al.}(2020)\citenamefont {Keidel},
  \citenamefont {Hwang}, \citenamefont {Trauzettel}, \citenamefont {Sothmann},\
  and\ \citenamefont {Burset}}]{keidel_PRR_2_022019}%
  \BibitemOpen
  \bibfield  {author} {\bibinfo {author} {\bibfnamefont {F.}~\bibnamefont
  {Keidel}}, \bibinfo {author} {\bibfnamefont {S.-Y.}\ \bibnamefont {Hwang}},
  \bibinfo {author} {\bibfnamefont {B.}~\bibnamefont {Trauzettel}}, \bibinfo
  {author} {\bibfnamefont {B.}~\bibnamefont {Sothmann}},\ and\ \bibinfo
  {author} {\bibfnamefont {P.}~\bibnamefont {Burset}},\ }\bibfield  {title}
  {\bibinfo {title} {On-demand thermoelectric generation of equal-spin cooper
  pairs},\ }\href {https://doi.org/10.1103/PhysRevResearch.2.022019} {\bibfield
   {journal} {\bibinfo  {journal} {Phys. Rev. Res.}\ }\textbf {\bibinfo
  {volume} {2}},\ \bibinfo {pages} {022019} (\bibinfo {year}
  {2020})}\BibitemShut {NoStop}%
\bibitem [{Note1()}]{Note1}%
  \BibitemOpen
  \bibinfo {note} {This form of Green's function is due to our particular
  choice of basis. Also, note that, this decomposition is different from that
  used in Phys.Rev.B 96, 155426(2017) and is similar to Eq. (S.39) of the
  Supplemental Material of Phys. Rev. Research 2, 022019(R)(2020)}\BibitemShut
  {NoStop}%
\bibitem [{\citenamefont {Cayao}\ and\ \citenamefont
  {Black-Schaffer}(2017)}]{cayao}%
  \BibitemOpen
  \bibfield  {author} {\bibinfo {author} {\bibfnamefont {J.}~\bibnamefont
  {Cayao}}\ and\ \bibinfo {author} {\bibfnamefont {A.~M.}\ \bibnamefont
  {Black-Schaffer}},\ }\bibfield  {title} {\bibinfo {title} {Odd-frequency
  superconducting pairing and subgap density of states at the edge of a
  two-dimensional topological insulator without magnetism},\ }\href
  {https://doi.org/10.1103/PhysRevB.96.155426} {\bibfield  {journal} {\bibinfo
  {journal} {Phys. Rev. B}\ }\textbf {\bibinfo {volume} {96}},\ \bibinfo
  {pages} {155426} (\bibinfo {year} {2017})}\BibitemShut {NoStop}%
\end{thebibliography}%
\end{document}


\title{Supplemental Material for ``Identification of odd-frequency superconducting pairing in Josephson junctions''}

\author{Subhajit Pal}
\affiliation{Indian Institute of Science Education \& Research Kolkata,
Mohanpur, Nadia - 741 246, 
West Bengal, India}

\author{Aabir Mukhopadhyay}
\affiliation{Indian Institute of Science Education \& Research Kolkata,
Mohanpur, Nadia - 741 246, 
West Bengal, India}

\author{Vivekananda Adak}
\affiliation{Department of Physics, Korea Advanced Institute of Science and Technology,
	Daejeon 34141, Korea}

\author{Sourin Das}
\affiliation{Indian Institute of Science Education \& Research Kolkata,
Mohanpur, Nadia - 741 246, 
West Bengal, India}
\pacs{74.50.+r,74.78.Na,85.25.-j,85.25.Cp,85.65.+h,75.50.Xx,85.80.Fi}

\maketitle

\onecolumngrid

\appendix

\renewcommand\appendixname{Supplemental Material}

\section{Derivation of scattering matrix}
\label{der_scat_mat}
We start with the Hamiltonian $H=H_{\text{HES}}+\sum_{i\in \{1,2\}}H_{P_i}+H_T^{P_i}$ as mentioned below Eq.~(5) in the main text. The equation of motion of an electron with spin $\alpha$, either within the HES or in the probes, can be written as
\begin{equation}
i\hbar\dot{\psi_{\alpha}}=[\psi_{\alpha},H].
\label{Eqn_of_motion}
\end{equation}
For stationary state problem $\dot{\psi}_{\alpha}=0$ (here $\alpha \in \{ \uparrow, \downarrow, \hat{n}_i \}$ as defined below Eq.~(5) in the main text). Thus, one can solve the following equation\cite{wad}
\begin{equation}
\lim_{\epsilon\rightarrow x_{i}}\int_{-\epsilon}^{\epsilon}[\psi_{\alpha},H]dx=0,
\label{Eqn_of_motion_2}
\end{equation}
to arrive at different matrix elements of $S^e$, as defined in Eq.~(6) in the main text. which turns out to be of the following form:
\begin{equation}
    \begin{pmatrix}
        S^{e}_{\uparrow\uparrow}    &S^{e}_{\uparrow\downarrow} &S^{e}_{\uparrow\hat{n}_i}  \\
        S^{e}_{\downarrow\uparrow}  &S^{e}_{\downarrow\downarrow}   &S^{e}_{\downarrow\hat{n}_i}   \\
        S^{e}_{\hat{n}_i\uparrow}   &S^{e}_{\hat{n}_i\downarrow}  &S^{e}_{\hat{n}_i\hat{n}_i}
    \end{pmatrix}_{x=x_i}  =
    \begin{pmatrix}
        \dfrac{4-t_{i}^2\cos(\theta_{i})}{4+t_{i}^2} &-\dfrac{t_{i}^2\sin(\theta_{i})}{4+t_{i}^2} &-\dfrac{4it_{i}\cos(\theta_{i}/2)}{4+t_{i}^2}   \\
        -\dfrac{t_{i}^2\sin(\theta_{i})}{4+t_{i}^2}  &\dfrac{4+t_{i}^2\cos(\theta_{i})}{4+t_{i}^2}    &-\dfrac{4it_{i}\sin(\theta_{i}/2)}{4+t_{i}^2}   \\
        -\dfrac{4it_{i}\cos(\theta_{i}/2)}{4+t_{i}^2}    &-\dfrac{4it_{i}\sin(\theta_{i}/2)}{4+t_{i}^2}   &\dfrac{4-t_{i}^2}{4+t_{i}^2}
    \end{pmatrix}.
    \label{scat_mat_e}
\end{equation}
Similar to Eq.~\eqref{Eqn_of_motion} and \eqref{Eqn_of_motion_2}, one can write the equation of motion for $\psi_{\alpha}^{\dagger}$. Particle-hole symmetry enables us to write the hole wave functions ($\psi_{\alpha}^h$) in terms of electron wave functions ($\psi_{\alpha}$) as
\begin{equation}
    \psi^{h}_{\alpha}=\eta(\alpha) \psi_{\alpha}^{\dagger},
\end{equation}
where we have considered $\eta(\alpha)=-1$ for $\alpha=\uparrow$ (within the HES) and $\eta=1$ otherwise. Note that, this convention is consistent with the particle-hole symmetry of the Nambu basis as in Eq.~(3) in the main text. In a similar spirit as in Eq.~(6) of the main text, one can define the scattering matrix of the hole, which turns out to be of the following form:
\begin{equation}
    \begin{pmatrix}
        S^{h}_{\uparrow\uparrow}    &S^{h}_{\uparrow\downarrow} &S^{h}_{\uparrow\hat{n}_i}  \\
        S^{h}_{\downarrow\uparrow}  &S^{h}_{\downarrow\downarrow}   &S^{h}_{\downarrow\hat{n}_i}   \\
        S^{h}_{\hat{n}_i\uparrow}   &S^{h}_{\hat{n}_i\downarrow}  &S^{h}_{\hat{n}_i\hat{n}_i}
    \end{pmatrix}_{x=x_i}  =
    \begin{pmatrix}
        \dfrac{4-t_{i}^2\cos(\theta_{i})}{4+t_{i}^2} &\dfrac{t_{i}^2\sin(\theta_{i})}{4+t_{i}^2} &-\dfrac{4it_{i}\cos(\theta_{i}/2)}{4+t_{i}^2}   \\
        \dfrac{t_{i}^2\sin(\theta_{i})}{4+t_{i}^2}  &\dfrac{4+t_{i}^2\cos(\theta_{i})}{4+t_{i}^2}    &\dfrac{4it_{i}\sin(\theta_{i}/2)}{4+t_{i}^2}   \\
        -\dfrac{4it_{i}\cos(\theta_{i}/2)}{4+t_{i}^2}    &\dfrac{4it_{i}\sin(\theta_{i}/2)}{4+t_{i}^2}   &\dfrac{4-t_{i}^2}{4+t_{i}^2}
    \end{pmatrix}.
    \label{scat_mat_h}
\end{equation}

\section{Derivation of Eq.~(7) in terms of pairing amplitudes}
\label{eqn_7_derv}
One can decompose the spin symmetry of retarded (advanced) Green's function $G^{r(a)}_{he}$ as\cite{keidel_PRR_2_022019}\footnote{This form of Green's function is due to our particular choice of basis. Also, note that, this decomposition is different from that used in Phys.Rev.B 96, 155426(2017) and is similar to Eq. (S.39) of the Supplemental Material of Phys. Rev. Research 2, 022019(R)(2020)}
\begin{equation}
    G_{he}^{r(a)}(x,x',\omega)=
    \begin{pmatrix}
        G_{he,\downarrow\uparrow}^{r(a)}(x,x',\omega)    &G_{he,\downarrow\downarrow}^{r(a)}(x,x',\omega) \\
        G_{he,\uparrow\uparrow}^{r(a)}(x,x',\omega)  &G_{he,\uparrow\downarrow}^{r(a)}(x,x',\omega)
    \end{pmatrix} =
    \sum_{j=0}^{3} f_j^{r(a)}(x,x',\omega) \sigma_j
    \label{Ghe_matrix_pair_amp}
\end{equation}
where $f^{r(a)}_{0}$, $f^{r(a)}_{1,2}$, $f^{r(a)}_{3}$ corresponds to spin-singlet, equal spin-triplet and mixed spin-triplet retarded (advanced) pairing amplitudes respectively. The relation between $f^{r}$ and $f^{a}$ is dictated by Fermi-Dirac statistics and is given by
\begin{equation}
    f_{0}^{r}(x,x',\omega)= f_{0}^{a}(x',x,-\omega),
    \qquad
    f_{j}^{r}(x,x',\omega)= -f_{j}^{a}(x',x,-\omega),
    \label{pairing_reln}
\end{equation}
where $j\in \{1,2,3 \}$. The off-diagonal elements in Eq.~\eqref{Ghe_matrix_pair_amp} are zero for our case.
Thus, we get
\begin{equation}
    G_{he,\downarrow\uparrow}^{r(a)}(x,x',\omega)=f_0^{r(a)}(x,x',\omega)+f_3^{r(a)}(x,x',\omega),
    \qquad
    G_{he,\uparrow\downarrow}^{r(a)}(x,x',\omega)=f_0^{r(a)}(x,x',\omega)-f_3^{r(a)}(x,x',\omega).
    \label{Greens_fun_pair}
\end{equation}
Using Eq.~\eqref{pairing_reln} and \eqref{Greens_fun_pair}, it can be easily shown that
\begin{equation}
    G_{he,\downarrow\uparrow}^{r}(x,x',\omega)=G_{he,\uparrow\downarrow}^{a}(x',x,-\omega),
    \qquad
    G_{he,\uparrow\downarrow}^{r}(x,x',\omega)=G_{he,\downarrow\uparrow}^{a}(x',x,-\omega).
    \label{G_ret_adv_rltn}
\end{equation}
The second identity in Eq.~\eqref{G_ret_adv_rltn} has been used to arrive at Eq.~(7) in the main text. Alternatively, one can use the particle-hole symmetry of the Green's function $\mathcal{P} G^{r}(x,x',\omega) \mathcal{P}^{-1} = -G^{r}(x,x',-\omega)$, where $\mathcal{P}=\sigma_y \tau_y \mathcal{K}$ with $\mathcal{K}$ being the operation of complex conjugation, to arrive at the identity
\begin{equation}
    G_{he,\uparrow\downarrow}^{r}(x,x',\omega) =G_{eh,\uparrow\downarrow}^{r*}(x,x',-\omega).
    \label{supple_equ_sym_1}
\end{equation}
Again, from identity $G^a(x,x',\omega)=[G^r(x',x,\omega)]^{\dagger}$, we have
\begin{equation}
    G_{eh,\uparrow\downarrow}^{r*}(x,x',-\omega)=G_{he,\downarrow\uparrow}^{a}(x',x,-\omega).
    \label{supple_equ_sym_2}
\end{equation}
Thus, by equating \eqref{supple_equ_sym_1} and \eqref{supple_equ_sym_2}, we arrive at the same result.

Now, the quantity we are interested in is $-|G^r_{he,\downarrow\uparrow}(x_2,x_1,\omega)|^2 +|G^a_{he,\downarrow\uparrow}(x_2,x_1,-\omega)|^2$. Using Eq.~\eqref{Greens_fun_pair} we can write
\begin{align}
   \mathcal{Q} &= -|G^r_{he,\downarrow\uparrow}(x_2,x_1,\omega)|^2 +|G^a_{he,\downarrow\uparrow}(x_2,x_1,-\omega)|^2    \nonumber   \\
   &= -|f_0^{r}(x_2,x_1,\omega)+f_3^{r}(x_2,x_1,\omega)|^2+|f_0^{a}(x_2,x_1,-\omega)+f_3^{a}(x_2,x_1,-\omega)|^2.
   \label{mathcalQ}
\end{align}
From this point onwards, we shall drop $(x_2,x_1,\pm\omega)$ in the parentheses, and it will be understood that advanced pairing amplitudes ($f^a$) correspond to $-\omega$. Also, it will be helpful to introduce the even ($f^{r,E}$) and odd parts ($f^{r,O}$) of the pairing amplitudes at this stage, defined as
\begin{equation}
    f^{r,E(O)}=\dfrac{f^r \pm f^a}{2}
    \qquad
    f^{r(a)}=f^{r,E} \pm f^{r,O}
    \label{even_odd_parts}
\end{equation}
Thus, from Eq.~\eqref{mathcalQ}, using Eq.~\eqref{even_odd_parts}, we get,
\begin{align}
    &-\mathcal{Q}   \nonumber \\
    &=(f_0^r+f_3^r)(f_0^{r*}+f_3^{r*}) -(f_0^a+f_3^a)(f_0^{a*}+f_3^{a*})    \nonumber   \\
    &= \left( f_0^{r} f_0^{r*}+f_3^{r} f_3^{r*} + f_0^{r}f_3^{r*}+f_0^{r*}f_3^{r}-(f_0^{a} f_0^{a*}+f_3^{a} f_3^{a*} + f_0^{a}f_3^{a*}+f_0^{a*}f_3^{a}) \right)  \nonumber  \\  
    &= 2 \left( (f_0^{r,O}+f_3^{r,O})(f_0^{r,E*}+f_3^{r,E*})+(f_0^{r,O*}+f_3^{r,O*})(f_0^{r,E}+f_3^{r,E}) \right)   \nonumber    \\
    &=4\, \text{Re} \left[ \left( f_0^{r,O}+f_3^{r,O} \right)  \left( f_0^{r,E*}+f_3^{r,E*}  \right) \right]    \nonumber\\
    &=4\, \text{Re}\left[\, \mathcal{O}[G^r_{he,\downarrow\uparrow}] \times \mathcal{E}[G^{r*}_{he,\downarrow\uparrow}] \,\right]
    \label{odd_even_product}
\end{align}
Thus, $-|G^r_{he,\downarrow\uparrow}(x_2,x_1,\omega)|^2 +|G^a_{he,\downarrow\uparrow}(x_2,x_1,-\omega)|^2 \propto \text{Re}\left[\, \mathcal{O}[G^r_{he,\downarrow\uparrow}(x_2,x_1,-\omega)] \times \mathcal{E}[G^{r*}_{he,\downarrow\uparrow}(x_2,x_1,-\omega)] \,\right]$ which has been used to derive Eq.~(7) in the main text. We can also use $\mathcal{A}(\kappa_{\downarrow\uparrow}^{21}) \propto -|G^{a}_{he,\downarrow\uparrow}(x_2,x_1,\omega)|^2 +|G^{a}_{he,\uparrow\downarrow}(x_1,x_2,\omega)|^2$ in Eq.~(7) of the main text instead of retarded Green's functions. In that case, we would have arrived at the same expression as Eq.~\eqref{odd_even_product} with $G^{r}_{he,\downarrow\uparrow}$ replaced by $G^{a}_{he,\downarrow\uparrow}$. Eq.~\eqref{odd_even_product} also justifies our statement that $\mathcal{A}(\kappa_{\downarrow\uparrow}^{21})$ is anti-symmetric with respect to $\omega$ and can be treated as a direct signature of odd-$\omega$ pairing when even-$\omega$ pairing is non-zero.

\section{Non-local transmission amplitudes in weak tunneling limit}
\label{non_local_diff_cond}
The wave functions in different regions of the system can be obtained by diagonalizing the Hamiltonians in Eq.~(3) and Eq.~(4) of the main text. By demanding the continuity of wave functions at $x=0,L$ and assuming that the incoming and outgoing plane wave amplitudes are related via Eqs.~\eqref{scat_mat_e} and \eqref{scat_mat_h} at $x=x_1$ and at $x=x_2$, one can arrive at the expressions for different non-local transmission amplitudes when an electron is injected through one of the probes, for arbitrary values of $\theta_1$ and $\theta_2$. The lowest order in $t_1$ and $t_2$, having non-zero terms in the Taylor series expansion of these quantities, is $O(t_1^{\,1}t_2^{\,1})$. Any lower order would physically mean that at least one of the probes is disconnected from the HES, resulting in the absence of non-local transmission amplitudes. In this lowest order, different transmission amplitudes read as
\begin{align}
    t_{ee}^{21}&= t_1 t_2 e^{i(k_e^{P_1}x_1-k_e^{P_2}x_2)} \left( -\dfrac{e^{i k_e(x_2-x_1) }\cos(\frac{\theta_1}{2})\cos(\frac{\theta_2}{2})}{1-e^{i(k_eL-k_hL-2\rho-\varphi_{21})}}  + \dfrac{e^{i k_e(x_1-x_2)}\sin(\frac{\theta_1}{2})\sin(\frac{\theta_2}{2})}{1-e^{-i(k_eL-k_hL-2\rho+\varphi_{21})}} \right),  \label{cc1}\\
    t_{he}^{21}&= t_1 t_2 e^{i(k_e^{P_1}x_1+k_h^{P_2}x_2)} \left( -\dfrac{e^{i(k_ex_1-k_hx_2-\rho)}\sin(\frac{\theta_1}{2})\cos(\frac{\theta_2}{2})}{1-e^{i(k_eL-k_hL-2\rho+\varphi_{21})}}  -\dfrac{e^{-i(k_ex_1-k_hx_2-\rho)}\cos(\frac{\theta_1}{2})\sin(\frac{\theta_2}{2})}{1-e^{-i(k_eL-k_hL-2\rho-\varphi_{21})}} \right), \\
    t_{ee}^{12}&= t_1 t_2 e^{i(k_e^{P_2}x_2-k_e^{P_1}x_1)} \left( \dfrac{e^{ik_e(x_1-x_2)}\cos(\frac{\theta_1}{2})\cos(\frac{\theta_2}{2})}{1-e^{-i(k_eL-k_hL-2\rho-\varphi_{21})}} - \dfrac{e^{ik_e(x_2-x_1)}\sin(\frac{\theta_1}{2})\sin(\frac{\theta_2}{2})}{1-e^{i(k_eL-k_hL-2\rho+\varphi_{21})}} \right),  \\
    t_{he}^{12}&= t_1 t_2 e^{i(k_e^{P_2}x_2+k_h^{P_1}x_1)} \left( -\dfrac{e^{-i(k_ex_2-k_hx_1-\rho)}\sin(\frac{\theta_1}{2})\cos(\frac{\theta_2}{2})}{1-e^{-i(k_eL-k_hL-2\rho-\varphi_{21})}} - \dfrac{e^{i(k_ex_2-k_hx_1-\rho)}\cos(\frac{\theta_1}{2})\sin(\frac{\theta_2}{2})}{1-e^{i(k_eL-k_hL-2\rho+\varphi_{21})}} \right),   \label{cc4}
\end{align}
where $\rho=\arccos(\omega/\Delta_0)$ (for $|\omega| \leq \Delta_{0})$ and 
$k_{e,h}=\mu\pm \omega$ are the wave-vectors in the metallic region while $k_{e,h}^{P_1,P_2}$ are the corresponding wave-vectors in the tunnelling probes $P_1$ and $P_2$. The quantities $t_{ee}^{21(12)}$ and $t_{he}^{21(12)}$ are electron and hole transmission amplitudes from $P_1(P_2)$ to $P_2(P_1)$.

Now, the expressions for different relevant free Green's functions within the normal region of a JJ based on HES are given as\cite{cayao}
\begin{align}
    G^{r,f}_{ee,\uparrow\uparrow}(x_2>x_1,\omega)&=\frac{1}{i}\frac{e^{i k_e(x_2-x_1)}}{1-e^{i(k_eL-k_hL-2\rho-\varphi_{21})}} =G^{a,f}_{ee,\uparrow\uparrow}(x_2>x_1,\omega),  \label{cc5}\\
    G^{r,f}_{ee,\uparrow\uparrow}(x_1<x_2,\omega) &=\frac{1}{i}\frac{-e^{ik_e(x_1-x_2)}}{1-e^{-i(k_eL-k_hL-2\rho-\varphi_{21})}} =G^{a,f}_{ee,\uparrow\uparrow}(x_1<x_2,\omega),   \\
    G^{r,f}_{ee,\downarrow\downarrow}(x_2>x_1,\omega)&=\frac{1}{i}\frac{-e^{i k_e(x_1-x_2)}}{1-e^{-i(k_eL-k_hL-2\rho+\varphi_{21})}} =G^{a,f}_{ee,\downarrow\downarrow}(x_2>x_1,\omega), \\
    G^{r,f}_{ee,\downarrow\downarrow}(x_1<x_2,\omega)&=\frac{1}{i}\frac{e^{ik_e(x_2-x_1)}}{1-e^{i(k_eL-k_hL-2\rho+\varphi_{21})}}= G^{a,f}_{ee,\downarrow\downarrow}(x_1<x_2,\omega),    \\
    G^{r,f}_{he,\downarrow\uparrow}(x,x',\omega) &=\frac{1}{i} \frac{-e^{-i(k_e x'-k_h x-\rho)}}{1-e^{-i(k_eL-k_hL-2\rho-\varphi_{21})}} = G^{a,f}_{he,\downarrow\uparrow}(x,x',\omega),  \\
    G^{r,f}_{he,\uparrow\downarrow}(x,x',\omega) &= \frac{1}{i} \frac{e^{i(k_e x'-k_h x-\rho)}}{1-e^{i(k_eL-k_hL-2\rho+\varphi_{21})}} =G^{a,f}_{he,\uparrow\downarrow}(x,x',\omega), \label{cc12}
\end{align}
($\{ x,x' \} \in \{ x_1,x_2 \}$). Now, one can use Eqs.~\eqref{cc5}-\eqref{cc12} to re-express the transmission amplitudes {in Eqs.~\eqref{cc1}-\eqref{cc4}} in terms of free Green's functions to arrive at the following expressions in the weak tunneling limit:
\begin{align}
	t_{ee}^{21}&=- i t_1t_2e^{i (k_e^{P_1}x_1-k_e^{P_2}x_2)} \left\{ G^{r(a),f}_{ee,\uparrow\uparrow}(x_2>x_1,\omega) \cos{(\theta_1/2)}\cos{(\theta_2/2)}+ G^{r(a),f}_{ee,\downarrow\downarrow}(x_2>x_1,\omega)\sin{(\theta_1/2)}\sin{(\theta_2/2)} \right\}   \label{t_ee_21}\, ,\\
	t_{he}^{21}&= i t_1t_2 e^{i(k_e^{P_1}x_1+k_h^{P_2}x_2)} \left\{ G^{r(a),f}_{he,\downarrow\uparrow}(x_2,x_1,\omega) \cos{(\theta_1/2)\sin{(\theta_2/2)}}-G^{r(a),f}_{he,\uparrow\downarrow}(x_2,x_1,\omega) \sin{(\theta_1/2)} \cos{(\theta_2/2)} \right\}   \label{t_he_21}\, ,\\
	t_{ee}^{12} &= - i t_1 t_2 e^{i(k_e^{P_2}x_2-k_e^{P_1}x_1)} \left\{ G^{r(a),f}_{ee,\uparrow\uparrow}(x_1<x_2,\omega) \cos{(\theta_1/2)} \cos{(\theta_2/2)} +G^{r(a),f}_{ee,\downarrow\downarrow}(x_1<x_2,\omega) \sin{(\theta_1/2)} \sin{(\theta_2/2)} \right\}    \label{t_ee_12}\, ,\\
	t_{he}^{12} &= i t_1 t_2 e^{i(k_e^{P_2}x_2+k_h^{P_1}x_1)} \left \{ G^{r(a),f}_{he,\downarrow\uparrow}(x_1,x_2,\omega) \sin{(\theta_1/2)} \cos{(\theta_2/2)} -G^{r(a),f}_{he,\uparrow\downarrow}(x_1,x_2,\omega) \cos{(\theta_1/2)} \sin{(\theta_2/2)} \right\}.   \label{t_he_12}\, 
\end{align}
Note that, TABLE.~I of the main text is consistent with the general expressions derived in Eq.~\eqref{t_ee_21}- \eqref{t_he_12}.

\begin{figure}[ht]
\centering{\includegraphics[width=0.9\textwidth]{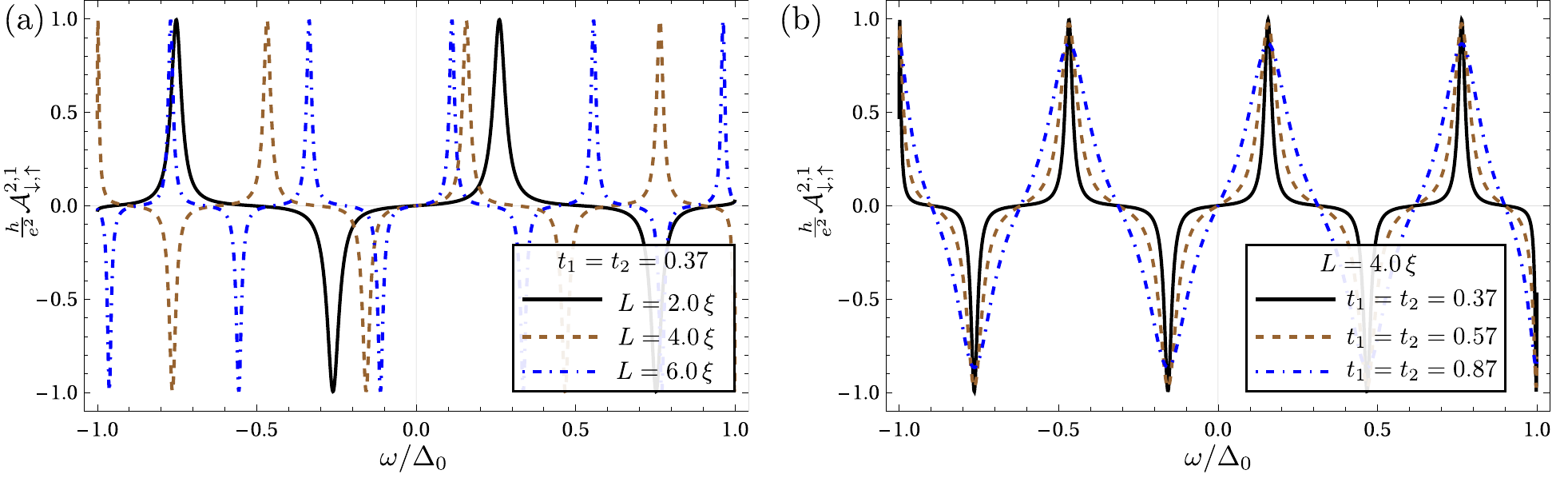}}
	\caption{\small \sl {Anti-symmetry in non-local differential conductance $\mathcal{A}^{2,1}_{\downarrow,\uparrow}$ as a function of $\omega$ for (a) different junction lengths $L$ and, (b) different tunneling strengths $t_1$ and $t_2$. Parameters are: $\mu=0$, $\varphi_{21}=\pi/2$, $\theta_1=0$, $\theta_2=\pi$.}}
	\label{comp_results}
\end{figure}

{\section{Dependence of $\mathcal{A}$ on system parameters}}
{The anti-symmetry property of $\mathcal{A}$, which is the main result of our paper, remains unaffected by changes in system parameters. However, the count of peaks and dips and their width may vary, as illustrated in Fig.~\ref{comp_results}. In Fig.~1, $\mathcal{A}$ is plotted as a function of $\omega$ for (a) different junction lengths $L$ while maintaining constant tunneling strengths $t_1$ and $t_2$, and (b) different tunneling strengths $t_1$ and $t_2$ while maintaining constant junction length $L$. The number of peaks/dips increases with increasing $L$ as depicted in Fig.~1(a), while in Fig.~1(b), the width of the peaks/dips increases with increasing tunneling strength. The position of the peaks/dips signifies the presence of Andreev bound states at those energies.}     

\bibliography{paper.bib}